\documentclass[twocolumn,showpacs,aps,pra]{revtex4}
\usepackage{amssymb}
\usepackage{amsfonts}
\usepackage{dcolumn}
\usepackage{amsmath}
\usepackage{graphicx}
\usepackage{psfrag}
\usepackage{epstopdf}
\newcommand{\be} {\begin{equation}}
\newcommand{\ee} {\end{equation}}

\begin{document}

\title{Robust quantum state transfer using tunable couplers}
\author{Eyob A. Sete}
\email{esete@ee.ucr.edu} \affiliation{Department of Electrical and
Computer Engineering, University of California, Riverside,
California 92521, USA}
\author{Eric Mlinar}
\affiliation{Department of Electrical and Computer Engineering,
University of California, Riverside, California 92521, USA}
\author{Alexander N. Korotkov}
\affiliation{Department of Electrical and Computer Engineering,
University of California, Riverside, California 92521, USA}
\date{\today}

\begin{abstract}
We analyze the transfer of a quantum state between two resonators
connected by a superconducting transmission line. Nearly perfect
state-transfer efficiency can be achieved by using adjustable
couplers and destructive interference to cancel the back-reflection into the  transmission line at the receiving coupler. We show that
the transfer protocol is robust to parameter variations affecting
the transmission amplitudes of the couplers. We also show that the
effects of Gaussian filtering, pulse-shape noise, and multiple
reflections on the transfer efficiency are insignificant. However,
the transfer protocol is very sensitive to frequency mismatch
between the two resonators. Moreover, the tunable coupler we
considered produces time-varying frequency detuning caused by the
changing coupling. This detuning requires an active frequency
compensation with an accuracy better than $90\%$ to yield the
transfer efficiency above $99\%$.
\end{abstract}
\maketitle

\section{Introduction}

The realization of quantum networks composed of many nodes requires
high-fidelity protocols that transfer quantum
states from site to site by using ``flying qubits''
\cite{Kim08,DiVincenzo-00}. The standard idea of the state transfer
between two nodes of a quantum network \cite{Cir97} assumes that the state of a
qubit is first encoded onto a photonic state at the emitting end,
after which the photon leaks out and propagates through a
transmission line to the receiving end, where its state is
transferred onto the second qubit. The importance of quantum state
transfer has stimulated significant research activity in optical
realizations of such protocols, e.g.,
\cite{Braunstein-98,Furusawa-98,Lloyd-01}, including trapping of
photon states in atomic ensembles \cite{Duan-01, Lukin-03, Chou-05,
Razavi-06}. Recent experimental demonstrations include the transfer
of an atomic state between two distant nodes \cite{Rem12} and the
transfer between an ion and a photon \cite{Bla13}.

An important idea for state transfer in the microwave domain is to
use tunable couplers between the quantum oscillators and the
transmission line \cite{Jah07,Kor11} (the idea is in general similar to the idea proposed in Ref.\ \cite{Cir97} for an optical system). In particular, this strategy
is natural for superconducting qubits, for which a variety of
tunable couplers have been demonstrated experimentally
\cite{Hime-06, Niskanen-07, Allman-10, Bialczak-11, Hoffman-11,
Yin13, Sri14, Wen14, Chen-14, Whittaker-14, Pierre-14} (these couplers are
important for many applications, e.g.,
\cite{Clarke-08,Gambetta-11,Sete-13,You-14}). Although there has been  rapid
progress in superconducting qubit technology, e.g.\
\cite{Barends-14, Chow-14, Weber-14, Sun-14, NEC-14, Stern-14,
Wal13, Gustavsson-13, Riste-13}, most of the experiments so
far are limited to a single chip or a single resonator in a dilution
refrigerator (an exception is \cite{Roch-14}). Implementing the quantum state transfer between remote
superconducting qubits, resonators, or even different refrigerators
using ``flying'' microwave qubits propagating through lossless
superconducting waveguides would significantly extend the capability of the technology (eventually permitting distributed quantum computing and quantum communications over extended distances using quantum repeaters). The essential ingredients of the transfer protocol proposed in Ref.\ \cite{Kor11} have already
been demonstrated experimentally. The emission of a proper
(exponentially increasing) waveform of a quantum signal has been demonstrated in
Ref.\ \cite{Sri14}, while the capture of such a waveform with 99.4\%
efficiency has been demonstrated in Ref.\ \cite{Wen14}. The
combination of these two procedures in one experiment would
demonstrate a complete quantum state transfer (more precisely, the
complete first half of the procedure of Ref.\ \cite{Kor11}). Note that Refs.\ \cite{Sri14} and \cite{Wen14} used different tunable couplers: a ``tunable mirror'' \cite{Yin13} between the resonator and the transmission line in Ref.\  \cite{Wen14} and a tunable coupling between the qubit and the resonator \cite{Hoffman-11} (which then rapidly decays into the transmission line) in Ref.\ \cite{Sri14}. However, this difference is insignificant for the transfer protocol of Ref.\ \cite{Kor11}. Another promising way to produce shaped photons is to use a modulated microwave drive to couple the superconducting qubit with the resonator
\cite{Pechal2014,Zeytinoglu2015} (see also Refs.\ \cite{Keller2004,Kolchin2008} for implementation of optical techniques for shaped photons).

In this work we extend the theoretical analysis of the state
transfer protocol proposed in Ref.\ \cite{Kor11}, focusing on its
robustness against various imperfections. In our protocol a quantum
state is transferred from the emitting resonator to the receiving
resonator through a transmission line (the state transfer using tunable coupling directly  between the qubit and the transmission line has also been considered in Ref.\ \cite{Kor11}, but we do not discuss it here). The procedure essentially
relies on the cancellation of back-reflection into the transmission
line via destructive interference at the receiving end, which is
achieved by modulation of the tunable couplers between the
resonators and the transmission line. (Note that the protocol is often discussed in terms of a ``time reversal'', following the terminology of Ref.\ \cite{Cir97}; however, we think that discussion in terms of a destructive interference is more appropriate.) In Ref.\ \cite{Kor11}, it was
shown that nearly perfect transfer efficiency can be achieved if
identical resonators and proper time-varying transmission amplitudes
of the two couplers are used. However, in obtaining this
high-efficiency state transfer, only ideal design parameters were
assumed. Also, various experimentally relevant effects, including
multiple reflections and frequency mismatch between the two
resonators, were not analyzed quantitatively.

In this paper we study in detail (mostly numerically) the effect of
various imperfections that affect the transmission amplitudes of the
couplers. In the simulations we focus on two values for the design
efficiency: 0.99 and 0.999. The value of 0.99 crudely corresponds to the current state of the art for the two-qubit quantum gate fidelities \cite{Barends-14} and threshold of some quantum codes \cite{Fowler2012}; we believe that the state transfer with 0.99 efficiency may already be interesting for practical purposes, while the value of 0.999 would be the next natural milestone for the experimental quantum state transfer. We find that the transfer protocol is
surprisingly robust to parameter variations, with a typical decrease
in the efficiency of less than 1\% for a 5\% variation of the design
parameters (the scaling is typically quadratic, so half of the variation produces a quarter of the effect). We also study
the effect of Gaussian filtering of the signals and find that it is
practically negligible. The addition of noise to the ideal waveforms
produces only a minor decrease in the transfer
efficiency. Numerical analysis of multiple reflections also shows
that the corresponding effect is not significant and can increase
the inefficiency by at most a factor of two. The analysis of the
effect of dissipative losses is quite simple and, as expected, shows that a high-efficiency state transfer requires a low-loss
transmission line and resonators with energy relaxation times
 much longer than duration of the procedure.

A major concern, however, is the effect of frequency mismatch
between the two resonators, since the destructive interference is
very sensitive to the frequency detuning. We consider two models: a
constant-in-time detuning and a time-dependent detuning due to
changing coupling. For the latter model we use the theory of the
coupler realized in Refs.\ \cite{Yin13,Wen14}; the frequency
variation due to the coupling modulation has been observed experimentally
\cite{Yin13}. Our results show that a high-efficiency state
transfer is impossible without an active compensation of the
frequency change; the accuracy of this compensation should be at
least within the 90\%-95\% range.

Although we assume that the state transfer is performed between
two superconducting resonators, using the tunable couplers
of Refs.\ \cite{Yin13,Wen14}, our analysis can also be applied to other setups, for example, schemes based on tunable couplers between the qubits and the transmission line or based on the tunable couplers between the qubits and the resonators \cite{Hoffman-11,Sri14,Pechal2014,Zeytinoglu2015}, which are then strongly coupled with the transmission line. Note that the frequency change compensation is done routinely in the coupler of Refs.\ \cite{Hoffman-11,Sri14},
thus giving a natural way to solve the problem of frequency mismatch. Similarly, the phase is naturally tunable in the coupler of Refs.\ \cite{Pechal2014,Zeytinoglu2015}.

The paper is organized in the following way. In Sec.\ \ref{model} we
discuss the ideal state transfer protocol, its mathematical model,
and the relation between classical transfer efficiency (which is
mostly used in this paper) and quantum state/process fidelity. In
Sec.\ \ref{imper} we analyze the decrease of the transfer efficiency
due to deviations from the design values of various parameters that
define the transmission amplitudes of the couplers. We also study
the effects of pulse-shape warping, Gaussian filtering, noise, and
dissipative losses. In Sec.\ \ref{mr} we analyze the effect of
multiple reflections of the back-reflected field on the transfer
efficiency. The effect of frequency mismatch between the two
resonators is discussed in Sec. \ref{fre-shift}. Finally, we
summarize the main results of the paper in Sec.\ \ref{con}. Appendix
A is devoted to the quantum theory of a beam splitter, which is used
to relate the efficiency of a classical state transfer to the
fidelity of a quantum state transfer. In Appendix B we discuss the
theory of the tunable coupler of Refs.\ \cite{Yin13,Wen14} and find
the frequency detuning caused by the coupling variation.

\section{Model and transfer protocol}\label{model}

\subsection{Model}

\begin{figure}[t]
\includegraphics[width=8cm]{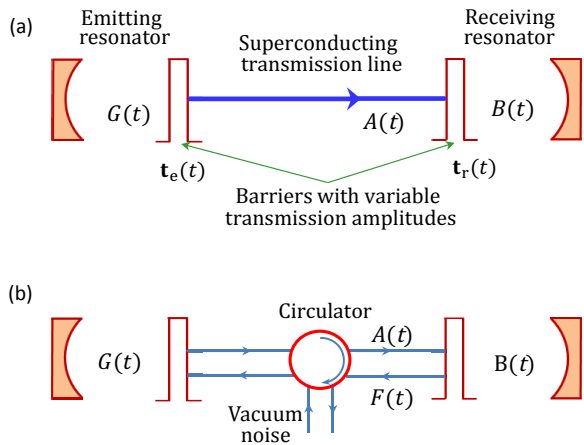}
  \caption{(a) The state transfer setup. An initial microwave field
amplitude $G(0)$ is transferred from the emitting resonator to the
receiving resonator via a transmission line. This is done using
variable couplers for both resonators, characterized by (effective)
transmission amplitudes ${\bf t}_{\rm e}(t)$ and ${\bf t}_{\rm
r}(t)$, and corresponding leakage rates $\kappa_{\rm e}(t)$ and
$\kappa_{\rm r}(t)$. Almost perfect transfer can be achieved when
the back-reflection of the propagating field $A(t)$ is cancelled by
arranging its destructive interference with the leaking part of the
field $B(t)$ in the receiving resonator. (b) A variant of the setup
that includes a circulator, which prevents multiple reflections of
the small back-reflected field $F(t)$. }
   \label{fig-model}
\end{figure}

\begin{figure}[t]
\includegraphics[width=8cm]{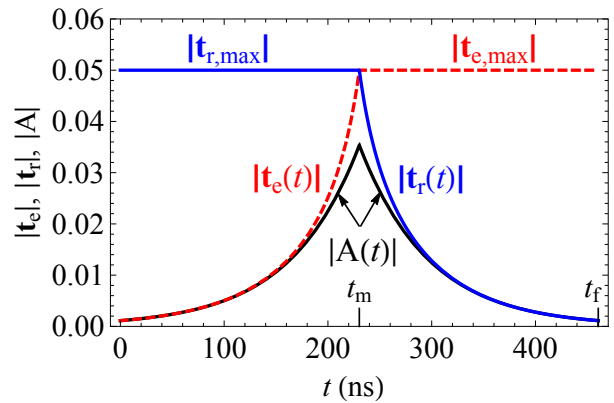}
  \caption{Time dependence (``pulse shapes'') of the absolute values of transmission
amplitudes ${\bf t}_{\rm e}(t)$ for the emitting coupler (red dashed
curve) and ${\bf t}_{\rm r}(t)$ for the receiving coupler (blue
solid curve). The amplitude ${\bf t}_{\rm e}(t)$ is kept constant at
the maximum level $\textbf{t}_{\rm e, max}$ after the mid-time
$t_{\rm m}$, while ${\bf t}_{\rm r}(t)$ is kept at the maximum
$\textbf{t}_{\rm r, max}$ during the first part of the procedure,
$t\leq t_{\rm m}$. The propagating field $A(t)$ first increases
exponentially and then decreases exponentially (black solid curve).
In simulations we typically use $|\textbf{t}_{\rm e,
max}|=|\textbf{t}_{\rm r, max}|=0.05$ for quarter-wavelength 6 GHz
resonators ($\tau_{\rm e}=\tau_{\rm r}=33$ ns); then the transfer
efficiency $\eta=0.999$ requires the procedure duration of $t_{\rm
f}=460$ ns.
   }
   \label{pulse-shapes}
\end{figure}

We consider the system illustrated in Fig.\ \ref{fig-model}(a). A
quantum state is being transferred from the emitting (left)
resonator into the initially empty receiving (right) resonator via
the transmission line. This is done by using time-varying couplings
(``tunable mirrors'') between the resonators and the transmission
line. The (effective) transmission amplitudes ${\bf t}_{\rm e}$ and
${\bf t}_{\rm r}$ for the emitting and receiving resonator couplers,
respectively, as a function of time $t$ are illustrated in Fig.\
\ref{pulse-shapes}. As discussed later, the main idea is to almost
cancel the back-reflection into the transmission line from the
receiving resonator by using destructive interference. Then the
field leaking from the emitting resonator is almost fully absorbed
into the receiving resonator. Ideally, we want the two resonators to
have equal frequencies, $\omega_{\rm e} =\omega_{\rm r}$; however,
in the formalism we will also consider slightly unequal resonator
frequencies $\omega_{\rm e}(t)$ and $\omega_{\rm r}(t)$. We assume
large quality factors $Q$ for both resonators by assuming $|{\bf
t}_{\rm e}(t)|\ll 1$ and $|{\bf t}_{\rm r}(t)|\ll 1$ (the maximum
value is crudely $|{\bf t}_{\rm e(r), max}| \sim 0.05$, leading to
$Q_{\rm min} \sim 10^3$ -- see later), so that we can use the
single-mode approximation. For simplicity, we assume a
dispersionless transmission line.

We will mostly analyze a classical field transfer between the two
resonators, with a straightforward relation to the quantum case,
discussed later. The notations $G(t)$ and $B(t)$ correspond to the
field amplitudes in the emitting and receiving resonators [see Fig.\
\ref{fig-model}(a)], while $A(t)$ describes the propagating field in
the transmission line. However, in contrast to the notations of
Ref.\ \cite{Kor11}, here we use dimensionless $G$ and $B$,
normalizing the field amplitudes \cite{Walls-book,Ger06} in such a
way that for classical (coherent) fields, $|G|^2$ and $|B|^2$ are
equal to the average number of photons in the resonators. Similarly,
the normalization of $A$ is chosen so that $|A|^2$ is the
number of propagating photons per second. Such normalizations for
resonators are more appropriate for the analysis of quantum
information. Also, with this normalization, the amplitudes will not
change with adiabatically-changing resonator frequency, in contrast
to the usual field  amplitudes.

In most of the analysis we assume (unless mentioned otherwise) that
the transmission line is either long or contains a circulator [Fig.\
\ref{fig-model}(b)], so that we can neglect the multiple reflections
of the small back-propagating field $F(t)$ (the effect of multiple
reflections will be considered in Sec.\ IV). We also assume that
there is no classical noise entering the emitting resonator from the
circulator (only vacuum noise).

With these assumptions and normalizations, the time dynamics of the
classical field amplitudes is described in the rotating frame by the
equations
    \begin{eqnarray}
    && \dot G = -i \Delta\omega_{\rm e} G -\frac{1}{2} \left( \kappa_{\rm e}+
    T_{1,\rm e}^{-1}\right) G,\label{lg}
    \\
    && \dot B =-i \Delta \omega_{\rm r} B - \frac{1}{2} \left( \kappa_{\rm r}+
    T_{1,\rm r}^{-1}\right)  B +
    \frac{\textbf{t}_{\rm r}}{|\textbf{t}_{\rm r}|}
    \sqrt{\kappa_{\rm r}}\, A,\label{le}
    \\
&&    A=\sqrt{\eta_{\rm tl}} \, \frac{\textbf{t}_{\rm
e}}{|\textbf{t}_{\rm e}|}
    \sqrt{\kappa_{\rm e}}\, G,
    \label{eq-A}\end{eqnarray}
where $\Delta \omega_{\rm e}=\omega_{\rm e}-\omega_0$ and $\Delta
\omega_{\rm r}=\omega_{\rm r}-\omega_0$ are small detunings
(possibly changing slowly with time) from the (arbitrary) rotating
frame frequency $\omega_0(t)$, the decay rates $\kappa_{\rm e}$ and
$\kappa_{\rm r}$ are due to leakage into the transmission line,
while additional losses are described by the energy relaxation times
$T_{1,\rm e}$ and $T_{1,\rm r}$ in the resonators  and imperfect
transfer efficiency $\eta_{\rm tl}$ of the transmission line. Note
that $A$ has the dimension of $1/\sqrt{\rm s}$ in contrast to the
dimensionless $G$ and $B$, so that the factors $\sqrt{\kappa_{\rm e
(r)}}$ restore the proper dimension.
 The
leakage rates are
    \be
    \kappa_{\rm e}(t)=\frac{|\tilde{\bf{t}}^{\rm in}_{\rm e}|^2}{\tau_{\rm
    rt,e}}\frac{R_{\rm e}}{R_{\rm tl}}=\frac{|\textbf{t}_{\rm e}|^2}{\tau_{\rm
    rt,e}}, \,\,\,
     \kappa_{\rm r}(t)=\frac{|\tilde{\bf{t}}^{\rm in}_{\rm r}|^2}{\tau_{\rm
    rt,r}}\frac{R_{\rm r}}{R_{\rm tl}}=\frac{|\textbf{t}_{\rm r}|^2}{\tau_{\rm
    rt,r}},
    \ee
where $\tilde{\bf{t}}_{\rm e}^{\rm in}$ and $\tilde{\bf{t}}_{\rm
r}^{\rm in}$ are the transmission amplitudes of the couplers (for a
wave incident from inside of the resonators), $\tau_{\rm rt,e}$ and
$\tau_{\rm rt,r}$ are the round-trip times in the resonators,
$R_{\rm e}$, $R_{\rm r}$, and $R_{\rm tl}$ are the wave impedances
of the resonators and the transmission line, while $\textbf{t}_{\rm
e}=\tilde{\bf{t}}_{\rm e}^{\rm in}\sqrt{R_{\rm e}/R_{\rm tl}}$ and
$\textbf{t}_{\rm r}=\tilde{\bf{t}}_{\rm r}^{\rm in}\sqrt{R_{\rm
r}/R_{\rm tl}}$ are the effective transmission amplitudes. Note that
the transmission amplitudes $\tilde{\bf t}$ depend on the wave
direction (from inside or outside of a resonator), while the
effective transmission amplitudes $\bf t$ do not. For convenience we
will be working with the effective transmission amplitudes
$\textbf{t}_{\rm e}$ and $\textbf{t}_{\rm r}$, so that we do not
need to worry about possibly unequal wave impedances. For
quarter-wavelength resonators $\tau_{\rm rt,e}\approx
\pi/\omega_{\rm e}\approx \pi/\omega_0$ and $\tau_{\rm rt,r}\approx
\pi/\omega_{\rm r}\approx \pi/\omega_0$, so the quality factors are
    \be
Q_{\rm e(r)}=\frac{\omega_{\rm e(r)}}{\kappa_{\rm
e(r)}}\approx \frac{\pi}{|\textbf{t}_{\rm e(r)}|^2}.
    \ee
 Note that the phase factors $\textbf{t}_{\rm r}/| \textbf{t}_{\rm r}|$ and
$\textbf{t}_{\rm e}/| \textbf{t}_{\rm e}|$ in Eqs.\ (\ref{le}) and
(\ref{eq-A}) may change in time because of changing coupling
\cite{Kor11,Yin13} (as discussed later in Sec.\ V$\,$B and Appendix B);
this is why these somewhat unusual factors cannot be neglected.
Strictly speaking, the last term in Eq.\ (\ref{le}) should also be
multiplied by $\sqrt{\omega_{\rm e}/\omega_{\rm r}}$; this is
because of different normalizations, related to different photon
energies $\hbar \omega_{\rm e}$ and $\hbar\omega_{\rm r}$ in the
resonators. However, we neglect this correction, assuming a
relatively small detuning. Note that the effective propagation time
along the transmission line is zero in Eqs.\
(\ref{lg})--(\ref{eq-A}) since we use appropriately shifted clocks
(here the assumption of a dispersionless transmission line is
necessary); however, the physical propagation time will be important
in the analysis of multiple reflections in Sec.\ IV. Also note that
to keep Eqs.\ (\ref{lg})--(\ref{eq-A}) reasonably simple, we defined the
phases of $B$ and $G$ to be somewhat different from the actual
phases of the standing waves in the resonators (see discussion in
Sec.\ II$\,$C).

Even though in Eqs.\ (\ref{lg})--(\ref{eq-A}) we use normalized
fields $G$, $B$, and $A$, which imply discussion in terms of the
photon number, below we will often use the energy terminology and
invoke the arguments of the energy conservation instead of the
photon number conservation. At least in the case without detuning
the two pictures are fully equivalent, but the energy language is
more intuitive, and thus preferable. This is why in the following we
will use the energy and photon number terminology interchangeably.


\subsection{Efficiency and fidelity}

We will characterize performance of the protocol via the transfer efficiency $\eta$, which is defined as the
ratio between the energy of the field (converted into the photon
number) in the receiving resonator at the end of the procedure,
$t=t_{\rm f}$, and the energy (photon number) at the initial time,
$t=0$, in the emitting resonator:
\begin{equation}\label{eff}
  \eta=\frac{|B(t_{\rm f})|^2}{|G(0)|^2}.
\end{equation}
We emphasize that in this definition we assume that only the
emitting resonator has initially a non-zero field.

As we discuss in this section, the classical efficiency $\eta$ is
sufficient to characterize the quantum transfer as well, so that the
quantum state and process fidelities derived below are directly
related to $\eta$ (this requires assumption of vacuum everywhere
except the initial state of the emitting resonator). The idea of the
conversion between the classical and quantum transfers is based on
the linearity of the process, and thus can be analyzed in
essentially the same way as the quantum optical theory of beam
splitters, discussed in Appendix A.

Let us focus on the case with the circulator [Fig.\ 1(b)] in the
absence of dissipative losses ($T_{1,\rm e}^{-1}=T_{1,\rm
r}^{-1}=0$, $\eta_{\rm tl}=1$). In general, there is a linear
input-output relation between the fields at $t=0$ and the fields at
$t=t_{\rm f}$. This relation is the same for the classical fields
and the corresponding quantum operators in the Heisenberg picture
(\cite{Yurke-84,Jah07}), so for simplicity we discuss the classical
fields. The relevant fields at $t=0$ are $G(0)$, $B(0)$, and the
(infinite number of) temporal modes propagating towards the emitting
resonator through the circulator; these modes can be described as
time-dependent field $V(t)$, where $t$ corresponds to the time, at which
the field arrives to the emitting resonator. Note that $B(0)$ and
$V(t)$ are assumed to be zero in our protocol; however, we need to
take them into account explicitly, because in the quantum language
they would correspond to operators, representing vacuum
noise (with the standard commutation relations). The fields at the
final time $t=t_{\rm f}$ are $B(t_{\rm f})$, $G(t_{\rm f})$, and the
collection of the outgoing back-reflected fields $F(t)$ for $0\leq
t\leq t_{\rm f}$ [see Fig.\ 1(b)]. Note that normalization of the
propagating fields $V(t)$ and $F(t)$ is similar to the normalization
of $A(t)$.

The input-output relation $\{ G(0), B(0), V(t)|_{0 \leq t\leq t_{\rm
f}}\} \mapsto \{ G(t_{\rm f}), B(t_{\rm f}), F(t)|_{0\leq t\leq
t_{\rm f}}\} $ is linear and unitary, physically because of the
conservation of the number of photons (energy). In particular,
    \be
    B(t_{\rm f})=\sqrt{\eta} \, e^{i\varphi_{\rm f}} G(0) + w_B B(0)+
\int_0^{t_{\rm f}} w_V(t) V(t)\, dt,
    \label{Bf-general}\ee
where $\eta$ is obviously given by Eq.\ (\ref{eff}), $\varphi_{\rm
f}$ is the phase shift between  $B(t_{\rm f})$ and $G(0)$, while
$w_B$ and $w_V(t)$ are some weight factors in this general linear
relation. These weight factors can be calculated by augmenting Eqs.\
(\ref{lg})--(\ref{eq-A}) to include $V(t)$ and $F(t)$, but we do not
really need them to find the quantum transfer fidelity if $B(0)$ and
$V(t)$ correspond to vacuum. Note that the unitarity of the
input-output transformation requires the relation
    \be
    \eta + |w_B|^2 + \int_0^{t_{\rm f}} |w_V(t)|^2 \, dt =1
    \label{unitarity}\ee
(sum of squared absolute values of elements in a row of a unitary matrix equals one), where we neglected the slight change in the normalization (discussed
above) in the case of time-varying detuning.

This picture of the input-output relations can in principle be
extended to include non-zero $T_{1,\rm e(r)}^{-1}$ and/or $\eta_{\rm
tl}\neq 1$; for that we would need to introduce additional noise
sources, which create additional terms in Eqs.\ (\ref{Bf-general})
and (\ref{unitarity}) similar to the terms from the noise $V$. Also,
if we consider the case without the circulator, the structure of
these equations remains similar, but the role of $V(t)$ is played by
the temporal modes of the initial field propagating in the
transmission line from the receiving to the emitting resonator
(since clocks are shifted along the transmission line, there is
formally no field ``stored'' in the transmission line, which
propagates from the emitting to the receiving resonator).

Using the framework of the linear input-output relation, Eq.\
(\ref{Bf-general}) derived for classical fields can also be used to
describe the quantum case. This can be done using the standard
quantum theory of beam splitters \cite{Ger06} (see Appendix A), by
viewing Eq.\ (\ref{Bf-general}) as the result of mixing the fields
$G(0)$, $B(0)$, and an infinite number of fields (temporal modes)
$V(t)$ with beam splitters to produce the proper linear combination.
Importantly, if $B(0)$ corresponds to vacuum and $V(t)$ also
corresponds to vacuum, then we can assume only one beam splitter
with the proper transfer amplitude $\sqrt{\eta}\, e^{i\varphi_{\rm
f}}$ for $G(0)\rightarrow B(t_{\rm f})$; this is because a linear
combination of several vacua is still the vacuum. Equivalently, the
resulting quantum state in the receiving resonator is equal to the
initial quantum state of the emitting resonator, subjected to the
phase shift $\varphi_{\rm f}$ and leakage (into vacuum) described by
the (classical) efficiency $\eta$. The same remains correct in the
presence of nonzero relaxation rates $T_{1,\rm e}^{-1}$ and
$T_{1,\rm r}^{-1}$ and imperfect $\eta_{\rm tl}$ if these processes
occur at zero effective temperature (involving only vacuum noise).

As shown in Appendix A, if the initial state in the emitting
resonator is $|\psi_{\rm in}\rangle=\sum_n \alpha_n |n\rangle$ in
the Fock space ($\sum_n |\alpha_n|^2=1$), then the final state of
the receiving resonator is represented by the density matrix, which
can be obtained from the state $|\psi_{\rm fin}\rangle =\sum_{n,k}
\alpha_{n+k} \sqrt{(n+k)!/n!k!} \, \eta^{n/2} (1-\eta)^{k/2} e^{i
(n+k)\varphi_{\rm f}}  |n\rangle|k\rangle_{\rm a}$  by tracing over
the ancillary state $|k\rangle_{\rm a}$ (this ancilla corresponds to
the second outgoing arm of the beam splitter). This gives the
density matrix $\rho_{\rm
fin}=\sum_{j,n,m}\alpha_{n+j}\alpha^{*}_{m+j}
\sqrt{(n+j)!(m+j)!} (j!\sqrt{n!m!})^{-1} \eta^{(n+m)/2} \\
(1-\eta)^je^{i(n-m)\varphi_{\rm f}})|n\rangle\langle m|$. The state
fidelity (overlap with the initial state) is then
\begin{align}
F_{\rm st}&=\sum_{j,n,m}\frac{\sqrt{(n+j)!(m+j)!}}{j!\sqrt{n!m!}} \,
\alpha_{n}^{*}\alpha_{m}\alpha_{n+j}\alpha^{*}_{m+j}
     \notag\\
&\hspace{0.5cm} \times \eta^{(n+m)/2}(1-\eta)^j
e^{i(n-m)\varphi_{\rm f}}.
    \label{F-st-gen}\end{align}
Note that the phase shift $\varphi_{\rm f}$ can easily be corrected
in an experiment (this correction is needed anyway for resonators,
which are significantly separated in space), and then the factor
$e^{i(n-m)\varphi_{\rm f}}$ in Eq.\ (\ref{F-st-gen}) can be removed.

The discussed quantum theory (at zero temperature, i.e., with only
vacuum noise) becomes very simple if we transfer a qubit state
$|\psi_{\rm in}\rangle=\alpha |0\rangle +\beta |1\rangle$. Then the
resulting state is
    \be
    |\psi_{\rm fin}\rangle = \alpha |0\rangle |0\rangle_{\rm a}
    +\beta e^{i\varphi_{\rm f}} (\sqrt{\eta}\, |1\rangle|0\rangle_{\rm a}+
    \sqrt{1-\eta}\,     |0\rangle|1\rangle_{\rm a}),
    \label{quantum-qubit}\ee
where the ancillary states $|1\rangle_{\rm a}$ and $|0\rangle_{\rm
a}$ indicate whether a photon was lost to the environment or not.
After tracing $ |\psi_{\rm fin}\rangle \langle \psi_{\rm fin}|$ over
the ancilla we obtain density matrix
    \be
    \rho_{\rm fin}=  \left(\begin{array}{cc} \eta |\beta |^2
& \sqrt{\eta} \, e^{i\varphi_{\rm f}} \alpha^* \beta  \\
\sqrt{\eta} \,  e^{-i\varphi_{\rm f}} \alpha \beta^* & |\alpha |^2 +|\beta |^2 (1-\eta)
    \end{array}\right) .
    \label{quantum-qubit-rho}\ee

Note that since a qubit state contains at most one excitation, the
essential dynamics occurs only in the single-photon subspace.
Therefore, it is fully equivalent to the dynamics of classical
fields (with field amplitudes replaced by probability amplitudes).
Thus, Eq.\ (\ref{quantum-qubit}) can be written directly, without
using the quantum beam splitter approach, which is necessary only
for multi-photon states.

In quantum computing the qubit state transfer (quantum channel) is
usually characterized by the quantum process fidelity $F_\chi$ or by
the average state fidelity $\overline{F}_{\rm st}$, which are
related as \cite{Nie02,Hor99} $1-F_\chi =(1-\overline{F}_{\rm
st})\times 3/2$. In order to calculate $F_\chi$, we calculate state fidelity $F_{\rm st}$ (overlap with initial state) and then average it over the Bloch sphere.  Neglecting the phase $\varphi_{\rm f}$, which can
be easily corrected in an experiment, from Eq.\
(\ref{quantum-qubit-rho}) we find $F_{\rm st}=|\alpha|^4+\eta |\beta|^4+|\alpha
\beta|^2 (1-\eta+2\sqrt{\eta})$, which also follows from Eq.\
(\ref{F-st-gen}). To average this fidelity over the Bloch sphere of
initial states, it is sufficient \cite{Nie02} (see also
\cite{Kea12}) to average it over only six states: $|0\rangle$,
$|1\rangle$, $(|0\rangle\pm|1\rangle)/\sqrt{2}$, and $(|0\rangle\pm
i|1\rangle)/\sqrt{2}$. This gives $\overline{F}_{\rm
st}=(3+\eta+2\sqrt{\eta})/6$, which can be converted into the
process fidelity
  \begin{equation}\label{pfn}
  F_{\chi}=\frac{1}{4}(1+\sqrt{\eta})^2.
  \end{equation}
This equation gives the relation between the classical energy
transfer efficiency $\eta$ which we use in this paper and the
process fidelity $F_\chi$ used in quantum computing. Note the
relation $1-F_\chi\approx (1-\eta)/2$ when $\eta \approx 1$. Also
note that a non-vacuum noise contribution (due to finite
temperature) always decreases $F_\chi$ (see Appendix A). If the
phase shift $\varphi_{\rm f}$ is included in the definition of
fidelity (assuming that $\varphi_{\rm f}$ is not corrected), then
Eq.\ (\ref{pfn}) becomes $F_{\chi}= (1+\eta+2\sqrt{\eta}\cos
\varphi_{\rm f})/4$.

Thus, in this section we have shown that the state and the process
fidelities of the quantum state transfer are determined by the
classical efficiency $\eta$ and experimentally correctable phase
shift $\varphi_{\rm f}$. This is why in the rest of the paper we
analyze the efficiency $\eta$ of essentially a classical state
transfer.

\subsection{Transfer procedure}

Now let us describe the transfer protocol, following Ref.\
\cite{Kor11} (this will be the second protocol out of two slightly
different procedures considered in Ref.\ \cite{Kor11}). Recall that
we consider normalized classical field amplitudes. The main idea of
achieving nearly perfect transfer is to use time-dependent
transmission amplitudes $\textbf{t}_{\rm e}$ and $\textbf{t}_{\rm
r}$ to arrange destructive interference between the field $A$
reflected from the receiving resonator and the part of field $B$
leaking through the coupler (see Fig.\ \ref{fig-model}). Thus, we
want the total back-reflected field $F(t)$ to nearly vanish:
$ F(t)\approx 0$, where
    \be
    F=\frac{\textbf{r}_{\rm r}^{\rm out}}
    {|\textbf{r}_{\rm r}|}\, A +
    \frac{\textbf{t}_{\rm r}}{|\textbf{t}_{\rm r}|}\,
    \sqrt{\kappa_{\rm r}} \,
    \frac{|\textbf{r}_{\rm r}|}{\textbf{r}_{\rm r}^{\rm in}} \,B,
    \label{F-form1}\ee
$\textbf{r}_{\rm r}^{\rm out}$ and $\textbf{r}_{\rm r}^{\rm in}$ are
the coupler reflection amplitudes from the outside and inside of the
receiving resonator, and $|\textbf{r}_{\rm r}|=|\textbf{r}_{\rm
r}^{\rm in}|=|\textbf{r}_{\rm r}^{\rm out}|$. Note that the
(effective) scattering matrix of the receiving resonator coupler is
$\left(\begin{array}{cc} \textbf{r}_{\rm r}^{\rm out}&
\textbf{t}_{\rm r}\\
\textbf{t}_{\rm r}&\textbf{r}_{\rm r}^{\rm in}\end{array}\right)$,
when looking from the transmission line. The formula (\ref{F-form1})
looks somewhat unusual for two reasons. First, in the single-mode
formalism of Eqs.\ (\ref{lg})--(\ref{eq-A}), the reflection
amplitude in Eq.\ (\ref{F-form1}) must be treated as having the
absolute value of 1; this is why we have the pure phase factor
$\textbf{r}_{\rm r}^{\rm out}/|\textbf{r}_{\rm r}|$. This is rather
counterintuitive and physically stems from the single-mode
approximation, which neglects the time delay due to the round-trip
propagation  in a resonator. It is easy to show that if the actual
amplitude $\textbf{r}_{\rm r}^{\rm out}$ were used for the
reflection $A\rightarrow F$, then solution of Eqs.\ (\ref{le}) and
(\ref{F-form1}) would lead to the energy non-conservation on the
order of $|\textbf{t}|^2$. Second, in our definition the phase of
the field $B$ corresponds to the standing wave component (near the
coupler) propagating away from the coupler [see Eq.\ (\ref{le})], so
the wave incident to the coupler is $B\, |\textbf{r}_{\rm
r}|/\textbf{r}_{\rm r}^{\rm in}$, thus explaining the phase factor
in the last term of Eq.\ (\ref{F-form1}). Actually, a better way
would be to define $B$ using the phase of the standing wave in the
resonator; this would replace the last term in Eq.\ (\ref{le}) with
$(\textbf{t}_{\rm r}/|\textbf{t}_{\rm r}|)\sqrt{\kappa_{\rm r}}\,  A
\sqrt{|\textbf{r}_{\rm r}|/\textbf{r}_{\rm r}^{\rm in}}$ and replace
the last term in Eq.\ (\ref{F-form1}) with $(\textbf{t}_{\rm
r}/|\textbf{t}_{\rm r}|) \sqrt{\kappa_{\rm r}}
\sqrt{|\textbf{r}_{\rm r}|/\textbf{r}_{\rm r}^{\rm in}} \,B$.
However, we do not use this better definition to keep a simpler form
of Eq.\ (\ref{le}).

Using the fact that $\textbf{t}^2_{\rm r}/\textbf{r}_{\rm r}^{\rm
in} \textbf{r}_{\rm r}^{\rm out}$ is necessarily real and negative
[since $\textbf{r}_{\rm r}^{\rm out}=-(\textbf{r}_{\rm r}^{\rm
in})^* \textbf{t}_{\rm r}/\textbf{t}_{\rm r}^*$ from unitarity], we
can rewrite Eq.\ (\ref{F-form1}) as
    \be
    F= \frac{\textbf{r}_{\rm r}^{\rm out}}{|\textbf{r}_{\rm r}|} \left( A - \frac{\textbf{t}^*_{\rm r}}{|\textbf{t}_{\rm r}|}\,\sqrt{\kappa_{\rm r}} \, B \right) .
    \label{F-form2}\ee
This form shows that if the phases of $\textbf{t}_{\rm r}$ and $A$
do not change in time and there is no detuning, then the two terms
in Eq.\ (\ref{F-form2}) have the same phase [because  $\arg
(B)=\arg(\textbf{t}_{\rm r}A)$  from Eq.\ (\ref{le})]. Therefore, for
the desired cancellation of the terms we need only the cancellation
of absolute values, i.e., a {\it one-parameter condition}.

For a non-zero field $B$, the exact back-reflection cancellation can
be achieved by varying in time the emitting coupling
$\textbf{t}_{\rm e}$ \cite{Jah07}, which determines $A$ in Eq.\
(\ref{F-form1}) or by varying the receiving coupling
$\textbf{t}_{\rm r}$ or by varying both of them with an appropriate
ratio \cite{Kor11}. At the very beginning of the procedure the exact
cancellation is impossible because $B(0)=0$, so there are two ways
to arrange an almost perfect state transfer. First, we can allow for
some loss during a start-up time $t_{\rm s}$ intended to create a
sufficient field $B$, and then maintain the exact cancellation of
the back-reflection at $t>t_{\rm s}$. Second, we can have a slightly
imperfect cancellation during the whole procedure. Both methods were
considered in Ref.\ \cite{Kor11}; in this paper we discuss only the
second method, which can be easily understood via an elegant
``pretend'' construction explained later.

Motivated by a simpler experimental realization, we divide our
protocol into two parts \cite{Kor11} (see Fig.\ \ref{pulse-shapes}).
During the first part of the procedure, we keep the receiving
coupler fixed at its maximum value $\textbf{t}_{\rm r, max}$, while
varying the emitting coupler to produce a specific form of $A(t)$
for an almost perfect cancellation. During the second part, we do
the opposite: we fix the emitting coupler at its maximum value
$\textbf{t}_{\rm e, max}$ and vary the receiving coupler. The
durations of the two parts are approximately equal.

The maximum available couplings between the resonators and
transmission line determine the timescales $\tau_{\rm e}$ and
$\tau_{\rm r}$ of the transfer procedure, which we define as the
inverse of the maximum leakage rates,
    \be
    \tau_{\rm e(r)} =\frac{1}{\kappa_{\rm e(r), max}}, \,\,\,
 \kappa_{\rm e(r), max}= \frac{|\textbf{t}_{\rm e(r), max}|^2} {\tau_{\rm rt,e(r)}}.
    \ee
The time $\tau_{\rm r}$ affects the buildup of the field in the
receiving resonator, while $\tau_{\rm e}$  determines the fastest
depopulation of the emitting resonator; we will call both $\tau_{\rm e}$
and $\tau_{\rm r}$  the buildup/leakage times.

Now let us discuss a particular construction \cite{Kor11} of the
procedure for nearly-perfect state transfer, assuming that the
complex phases of $\textbf{t}_{\rm e}$ and $\textbf{t}_{\rm r}$ are
constant in time, there is no detuning, $\omega_{\rm e}=\omega_{\rm
r}=\omega_0$, and there is no dissipative loss, $T_{1,\rm
e}^{-1}=T_{1,\rm r}^{-1}=0$, $\eta_{\rm tl}=1$. (For the
experimental coupler discussed in Appendix B, $\textbf{t}_{\rm e}$
and $\textbf{t}_{\rm r}$ are mostly imaginary, but also have a
significant real component.) As mentioned above, during the first
part of the procedure, the receiving resonator is maximally coupled,
$\textbf{t}_{\rm r}(t)=\textbf{t}_{\rm r, max}$, with this value
being determined by experimental limitations. Then a complete
cancellation of the back-reflection, $F=0$,  would be possible if
$A(t)=A_0 \exp (t/2\tau_{\rm r})$ and $B(t)=B_0 \exp (t/2\tau_{\rm
r})$ with $B_0=\sqrt{\tau_{\rm r}}\, A_0 \textbf{t}_{\rm r,
max}/|\textbf{t}_{\rm r, max}|$. This is simple to see from Eqs.\
(\ref{le}) and (\ref{F-form2}), and even simpler to see using the
time reversal symmetry: the absence of the back-reflection will then
correspond to a leaking resonator without an incident field. This is
why in the reversed-time picture $B\propto \exp(-t/2\tau_{\rm r})$,
and therefore in the forward-time picture $B\propto \exp(t/2\tau_{\rm r})$;
the same argument applies to $A$.

Thus, we wish to generate an exponentially increasing transmitted
field
    \be
    A(t) = A_0 \exp (t/2\tau_{\rm r}), \,\,\, 0\le t\le t_{\rm m},
    \label{A(t)-1}\ee
during the first half of the procedure (until the mid-time $t_{\rm
m}$) by increasing the emitting coupling $\textbf{t}_{\rm e}(t)$.
This would provide the perfect cancellation of reflection if
$B(0)=B_0$ (as in the above example), while in the actual case when
$B(0)=0$ we can still use the waveform (\ref{A(t)-1}), just
``pretending'' that $B(0)=B_0$. It is easy to see that this
provides an almost perfect cancellation. Let us view
the initially empty resonator as a linear combination: $B(0)=B_0-B_0$. Then
due to linearity of the evolution, the part $B_0$ will lead to
perfect cancellation as in the above example, while the part $-B_0$
will leak through the coupler and will be lost. If $-B_0$ is fully
lost during a sufficiently long procedure, then the corresponding
contribution to the inefficiency (mostly from the initial part of
the procedure) is $1-\eta_{\rm r}=|B_0/G(0)|^2$. In particular,
for a symmetric procedure ($\tau_{\rm e}=\tau_{\rm r}=\tau$, $t_{\rm
m}=t_{\rm f}/2$) approximately one half of the energy will be
transmitted during the first half of the procedure, $|B(t_{\rm
m})|^2\approx |G(0)|^2/2$; then $|B_0|^2\approx \exp(-t_{\rm
m}/\tau) \, |G(0)|^2/2$, and therefore the inefficiency contribution
is $1-\eta_{\rm r}\approx \exp(-t_{\rm m}/\tau)/2$. As we see,
the {\it inefficiency decreases exponentially} with the procedure
duration.

At time $t_{\rm m}$ the increasing emitting coupling
$\textbf{t}_{\rm e}$ reaches its maximum value $\textbf{t}_{\rm e,
max}$ (determined by experimental limitations), and after that we
can continue cancellation of the back-reflection (\ref{F-form2}) by
decreasing the receiving coupling $\textbf{t}_{\rm r}(t)$, while
keeping emitting coupling at $\textbf{t}_{\rm e, max}$. Then the
transmitted field $A(t)$ will become exponentially decreasing,
    \be
    A(t) = A_0 \exp (t_{\rm m}/2\tau_{\rm r})
    \exp[-(t-t_{\rm m})/2\tau_{\rm e}], \,\,\,
    t_{\rm m}\le t\le  t_{\rm f},
    \label{A(t)-2}\ee
and $\textbf{t}_{\rm r}$ should be varied correspondingly, so that
$\kappa_{\rm r}(t)=|A(t)|^2/|B(t)|^2$. As mentioned above, the phase
conditions for the destructive interference are satisfied
automatically in the absence of detuning and for fixed complex
phases of $\textbf{t}_{\rm e}(t)$ and $\textbf{t}_{\rm r}(t)$. The
procedure is stopped at time $t_{\rm f}$, after which
$\textbf{t}_{\rm r}(t)=0$, so that the receiving resonator field
$B(t_{\rm f})$ no longer changes. When the procedure is stopped at
time $t_{\rm f}$, there is still some field $G(t_{\rm f})$ remaining
in the emitting resonator. This leads to the inefficiency
contribution $1-\eta_{\rm e}=|G(t_{\rm f})/G(0)|^2$. Again
assuming a symmetric procedure ($\tau_{\rm e}=\tau_{\rm r}=\tau$,
$t_{\rm f}=2t_{\rm m}$), we can use $|B(t_{\rm m})|^2\approx
|B(0)|^2/2$; then $|B(t_{\rm f})|^2\approx \exp(-t_{\rm m}/\tau)
|B(0)|^2/2$  and therefore $1-\eta_{\rm e}=\exp(-t_{\rm m}/\tau
)/2$. Combining the two (equal) contributions to the inefficiency,
we obtain \cite{Kor11}
    \be
    1-\eta \approx \exp (-t_{\rm f}/2\tau) .
    \label{ineff-1}\ee
The numerical accuracy of this formula is very high when $t_{\rm
f}\agt 10 \tau$.

Now let us derive the time dependence of the couplings
$\textbf{t}_{\rm e}(t)$  and $\textbf{t}_{\rm r}(t)$ needed for this
almost perfect state transfer (we assume that $\tau_{\rm e}$ and
$\tau_{\rm r}$ can in general be different). Again, the idea of the
construction is to arrange {\it exact cancelation of the
back-reflection if there were an initial field $B_0$ in
the receiving resonator} (with proper phase). In this hypothetical
``pretend'' scenario the evolution of the receiving resonator field
$\tilde B(t)$ is slightly different from $B(t)$ in the actual case
[$B(0)=0$, $\tilde B(0)=B_0$], while the fields $G(t)$ and $A(t)$ do
not change. Thus, we consider the easy-to-analyze ideal ``pretend'' scenario
$\tilde B(t)$ and then relate it to the actual evolution $B(t)$.
Note that the transmitted field $A(t)$ is given by Eqs.\ (\ref{A(t)-1}) and (\ref{A(t)-2}): it is exponentially increasing until $t_{\rm m}$ and exponentially decreasing after  $t_{\rm m}$. Also note that our procedure does not involve optimization: the only parameter, which can be varied, is the duration of the procedure, which is determined by the desired efficiency (the only formal optimization will be a symmetric choice of $t_{\rm m}$).

In the first part of the procedure, $t\leq t_{\rm m}$, the receiving
coupling is at its maximum, $\textbf{t}_{\rm r}(t)=\textbf{t}_{\rm
r, max}$, and the emitting coupling can be found as $\textbf{t}_{\rm
e}(t)=\textbf{t}_{\rm e,max} \sqrt{\tau_{\rm e}} \, |A/G|$ (recall
that phase conditions are fixed). Here $A(t)$ is given by Eq.\
(\ref{A(t)-1}) and $|G(t)|$ can be found from energy conservation in
the ``pretend'' scenario: $|G(t)|^2+|B_0 \exp(t/2\tau_{\rm
r})|^2=|G(0)|^2+|B_0|^2$. Using the relation
$|B_0/A_0|=\sqrt{\tau_{\rm r}}$, we find
    \be
    \textbf{t}_{\rm e}(t)= \textbf{t}_{\rm e,max}\sqrt{\frac{\tau_{\rm e}}{\tau_{\rm r}}} \, \frac{\exp(t/2\tau_{\rm r})}{\sqrt{|G(0)/B_0|^2+1-\exp(t/\tau_{\rm r})}}.
    \label{t-e-1}\ee
Here $|B_0|$ is an arbitrary parameter (related to an arbitrary
$|A_0|$), which affects the efficiency and duration of the
procedure. The corresponding $G(t)$ and $B(t)$ evolutions are
    \begin{eqnarray}
&&    G(t)= G(0)\sqrt{1-|B_0/G(0)|^2 [\exp(t/\tau_{\rm r})-1]},
\qquad
    \\
&& B(t) = B_0 [\exp(t/2\tau_{\rm r})-\exp(-t/2\tau_{\rm r}) ].
    \label{B(t)-1}\end{eqnarray}
Note that in the ``pretend'' scenario $\tilde
B(t)=B_0\exp(t/2\tau_{\rm r})$, while actually $B(t)=\tilde B
(t)-B_0\exp(-t/2\tau_{\rm r})$, where the second term describes the
decay of the compensating initial field $-B_0$.
 The phase of $B_0$ is determined by the phases of the transmission
amplitudes, $\arg(B_0)=\arg[\textbf{t}_{\rm e,max}\textbf{t}_{\rm
r,max}G(0)]$.

Since $|B_0|$ is related to the mid-time $t_{\rm m}$ via the
condition  $\textbf{t}_{\rm e}(t_{\rm m})= \textbf{t}_{\rm e,max}$,
it is convenient to rewrite Eq.\ (\ref{t-e-1}) in terms of $t_{\rm
m}$. Thus, the resonator couplings during the first part of the
procedure should be \cite{Kor11}
    \begin{align}
&  \textbf{t}_{\rm e}(t)=\frac{\textbf{t}_{\rm e,
max}\sqrt{\tau_{\rm e}/\tau_{\rm r}}}{\sqrt{( 1+\tau_{\rm
e}/\tau_{\rm r}) \exp [(t_{\rm m}-t)/\tau_{\rm r}]-1}} ,
     \label{tem}\\
    & \textbf{t}_{\rm r}(t)=\textbf{t}_{\rm r, max}, \,\,\,\,\,\,
    0\leq t\leq t_{\rm m} .
    \end{align}
Note that the increase of $\textbf{t}_{\rm e}(t)$ is slightly faster
than exponential.

To derive the required $\textbf{t}_{\rm r}(t)$ during the second
part of the procedure, $t \geq t_{\rm m}$, we can use the time
reversal of the ``pretend'' scenario. It will then describe a
perfect field absorption by the emitting resonator; therefore,
$\textbf{t}_{\rm r}(t)$ in the reversed (and shifted) time should
obey the same Eq.\ (\ref{t-e-1}), but with exchanged indices
(e$\leftrightarrow$r) and $|G(0)/B_0|$ replaced with $|\tilde
B(t_{\rm f})/G(t_{\rm f})|$. Then by using the condition
$\textbf{t}_{\rm r}(t_{\rm m})= \textbf{t}_{\rm r,max}$ we
immediately derive the formula similar to Eq.\ (\ref{tem}),
    \begin{align}
&  \textbf{t}_{\rm r}(t)=\frac{\textbf{t}_{\rm r,
max}\sqrt{\tau_{\rm r}/\tau_{\rm e}}}{\sqrt{( 1+\tau_{\rm
r}/\tau_{\rm e}) \exp [(t-t_{\rm m})/\tau_{\rm e}]-1}} ,
     \label{trec}\\
    & \textbf{t}_{\rm e}(t)=\textbf{t}_{\rm e, max}, \,\,\,\,\,\,
    t_{\rm m} \leq t\leq t_{\rm f} .
    \label{trec-2}\end{align}
It is also easy to derive Eq.\ (\ref{trec}) as $\textbf{t}_{\rm
r}(t)=\textbf{t}_{\rm r,max} \sqrt{\tau_{\rm r}} \, |A/\tilde B|$,
with $A(t)$ given by Eq.\ (\ref{A(t)-2}) and $|\tilde B
(t)|^2=|G(0)|^2+|B_0|^2-|G(t)|^2$ given by the energy conservation,
where $|G(t)|=\sqrt{\tau_{\rm e}}\,|A(t)|$.

The contribution to the inefficiency due to imperfect reflection
(mostly during the initial part of the procedure) is $1-\eta_{\rm
r}\approx |B_0/G(0)|^2$ since the reflected field is the leaking
initial field $-B_0$ and it is almost fully leaked during the
procedure. Comparing Eqs.\ (\ref{t-e-1}) and (\ref{tem}), we find
$|B_0/G(0)|^2 \approx \exp (-t_{\rm m}/\tau_{\rm r}) \,\tau_{\rm r}/
(\tau_{\rm e}+\tau_{\rm r})$ assuming $\exp (-t_{\rm m}/\tau_{\rm
r})\ll 1$. The contribution to the inefficiency due to the
untransmitted field left in the emitting resonator at the end of
procedure is $1-\eta_{\rm e}= |G(t_{\rm f})/G(0)|^2=(\tau_{\rm
e}/\tau_{\rm r})\, |B_0/G(0)|^2 \exp (t_{\rm m}/\tau_{\rm r}) \exp
[-(t_{\rm f}-t_{\rm m})/\tau_{\rm e}]$, where we used relation
$|G(t_{\rm f})|^2=\tau_{\rm e}|A(t_{\rm f})|^2$. Using the above
formula for $|B_0/G(0)|^2$ we obtain $1-\eta_{\rm e} \approx
\exp [-(t_{\rm f}-t_{\rm m})/\tau_{\rm e}] \,\tau_{\rm e}/
(\tau_{\rm e}+\tau_{\rm r})$. Combining both contributions to the
inefficiency we find \cite{Kor11}
    \be
    1-\eta \approx \frac{\tau_{\rm r}\exp (-t_{\rm m}/\tau_{\rm r}) +
    \tau_{\rm e}\exp [-(t_{\rm f}-t_{\rm m})/\tau_{\rm r}]}
    {\tau_{\rm e}+\tau_{\rm r}}.
    \ee
Minimization of this inefficiency over $t_{\rm m}$ for a fixed total
duration $t_{\rm f}$ gives the condition
    \be
    t_{\rm m}/\tau_{\rm r} =(t_{\rm f}-t_{\rm m})/\tau_{\rm e}
    \ee
and the final result for the inefficiency \cite{Kor11},
    \be
    1-\eta \approx \exp \left( - \frac{t_{\rm f}}
    {\tau_{\rm e}+\tau_{\rm r}} \right) ,
    \label{ineff-2}\ee
which generalizes Eq.\ (\ref{ineff-1}).

The required ON/OFF ratios for the couplers can be found from Eqs.\
(\ref{tem}) and (\ref{trec}),
    \begin{eqnarray}
&&    \frac{\textbf{t}_{\rm e,max}}{\textbf{t}_{\rm e}(0)}\approx
    \sqrt{\frac{\tau_{\rm e}+\tau_{\rm r}}{\tau_{\rm e}}
    \exp \left( \frac{t_{\rm m}}{\tau_{\rm     r}}\right)},
     \\
&&     \frac{\textbf{t}_{\rm r,max}}{\textbf{t}_{\rm r}(t_{\rm f})}
     \approx \sqrt{\frac{\tau_{\rm e}+\tau_{\rm r}}{\tau_{\rm r}}
     \exp \left( \frac{t_{\rm f}-t_{\rm m}}{\tau_{\rm  r}}\right)
     },
     \end{eqnarray}
which in the optimized case corresponding to Eq.\ (\ref{ineff-2})
become
    \be
\frac{\textbf{t}_{\rm e,max}}{\textbf{t}_{\rm e}(0)}\approx
    \sqrt{\frac{1+\tau_{\rm r}/\tau_{\rm e}}{1-\eta}}, \,\,\,
\frac{\textbf{t}_{\rm r,max}}{\textbf{t}_{\rm r}(t_{\rm f})}\approx
    \sqrt{\frac{1+\tau_{\rm e}/\tau_{\rm r}}{1-\eta}}.
       \label{On-Off}\ee

Note that using two tunable couplers is crucial for our protocol. If only one tunable coupler is used as in Ref.\ \cite{Jah07}, then the procedure becomes much longer and requires a much larger ON/OFF ratio. Assuming a fixed receiving coupling, we can still use Eqs.\ (\ref{t-e-1})--(\ref{B(t)-1}) for the analysis and obtain the following result. If the coupling of the emitting resonator is limited by a maximum value $\kappa_{\rm max}$ of the leakage rate, then the shortest duration of the procedure with efficiency $\eta$ is $t_{\rm f}=LN/[\kappa_{\rm max}(1-\eta)]$, where
$LN \approx \ln \frac{ e \, \ln [(e/(1-\eta)]}{1-\eta}$. For typical values of $\eta$ we get $LN\approx 3+\ln [1/(1-\eta)]$, and therefore the shortest duration for a procedure with one tunable coupler is $t_{\rm f}\approx (1-\eta)^{-1} \kappa_{\rm max}^{-1} \, \{ 3+\ln [1/(1-\eta)] \}$. This is more than a factor $(1-\eta)^{-1}/2$ longer than the duration $t_{\rm f} =2\, \kappa_{\rm max}^{-1}  \ln [1/(1-\eta)]$ of our procedure with two tunable couplers [see Eq.\ (\ref{ineff-1})]. The optimum (fixed) receiving coupling is $\kappa_{\rm r}=(1-\eta)\kappa_{\rm max}/(1+1/LN)$, which makes clear why the procedure is so long. The corresponding ON/OFF ratio for the emitting coupler is $\textbf{t}_{\rm e,max}/\textbf{t}_{\rm e}(0)=\sqrt{\kappa_{\rm max}/\kappa_{\rm min}}= (1-\eta )^{-1} \sqrt{LN/(1-2/LN)}\approx (1-\eta )^{-1} \sqrt{ 3+\ln [1/(1-\eta)]}$. This is more than a factor $(1-\eta)^{-1/2}$ larger than what is needed for our procedure [see Eq.\ (\ref{On-Off})].

Note that we use the exponentially increasing and then exponentially decreasing transmitted field $A(t)$ [Eqs.\ (\ref{A(t)-1}) and (\ref{A(t)-2})] because we wish to vary only one coupling in each half of the procedure and to minimize the duration of the procedure. In general, any ``reasonable'' shape $A(t)$ can be used in our procedure. Assuming for simplicity a real positive $A(t)$, we see that a ``reasonable'' $A(t)$ should satisfy the inequality $A^2(t)\leq \kappa_{\rm e, max}[|G(0)|^2-\int_0^t A^2 (t')\, dt']$, so that it can be produced by using $\kappa_{\rm e}(t)=A^2(t)/[|G(0)|^2-\int_0^t A^2 (t')\, dt']$ without exceeding the maximum emitting coupling  $\kappa_{\rm e, max}$. We also assume that a ``reasonable'' $A(t)$ does not increase too fast, $dA(t)/dt \leq (\kappa_{\rm r, max}/2) A(t)$, or at least satisfies a weaker inequality $A(t)\leq \sqrt{\kappa_{\rm r, max}}\sqrt{\kappa_{\rm r, max}^{-1}A^2(0) +\int_0^t A^2(t')\,dt'}$. In this case we can apply the ``pretend'' method, which gives  $\kappa_{\rm r}(t)=A^2(t)/[\kappa_{\rm r, max}^{-1}A^2(0)+\int_0^t A^2(t')\,dt']$, not exceeding the maximum receiving coupling $\kappa_{\rm r, max}$. This leads to the inefficiency contribution $1-\eta_{\rm e}=1-\int_0^{t_{\rm f}} A^2 (t')\, dt'/|G(0)|^2$ due to the untransmitted field and inefficiency contribution $1-\eta_{\rm r}= \kappa_{\rm r, max}^{-1}A^2(0)/|G(0)|^2$ due to the back-reflection. We see that for high efficiency we need a small $A(t)$ at the beginning and at the end of the procedure. Even though we do not have a rigorous proof, it is intuitively obvious that our procedure considered in this section is optimal (or nearly optimal) for minimizing the duration of the protocol for a fixed efficiency and fixed maximum couplings (see also the proof of optimality for a similar, but single-sided procedure in Ref.\ \cite{Jah07}). We think that it is most natural to design an  experiment exactly as described in this section [using Eqs.\ (\ref{A(t)-1}) and (\ref{A(t)-2}) and varying only one coupling at a time]; however, a minor or moderate time-dependent tuning of the other coupling (which is assumed to be fixed in our protocol) can be useful in experimental optimization of the procedure.

In this section, we considered the ideal transfer protocol, assuming
that the transmission amplitudes are given exactly by Eqs.\
\eqref{tem}--\eqref{trec-2}, and also assuming equal resonator
frequencies, fixed phases of the transmission amplitudes, and
absence of extra loss ($T_{1,\rm e}^{-1}=T_{1,\rm r}^{-1}=0$,
$\eta_{\rm tl}=1$). In the following sections we will discuss the
effect of various imperfections on the efficiency of the transfer
protocol.

\section{Imperfect pulse shapes}\label{imper}

The high efficiency of the state transfer analyzed in the previous
section relies on precise calibration and control of experimental
parameters, so that the needed pulse shapes
\eqref{tem}--\eqref{trec-2} for the transmission amplitudes
$\textbf{t}_{\rm e}(t)$ and $\textbf{t}_{\rm r}(t)$ are accurately
implemented. However, in a real experiment there will always be some
imperfections in the pulse shapes. In this section we analyze the
robustness of the transfer efficiency to the pulse shape
imperfections, still assuming fixed phases and the absence of
detuning and dissipative loss. In particular, we will vary several
parameters used in the pulse shapes \eqref{tem}--\eqref{trec-2}: the
maximum transmission amplitudes $|\textbf{t}_{\rm e(r),max}|$, the
buildup/leakage times $\tau_{\rm e(r)}$, and the mid-time $t_{\rm
m}$. By varying these parameters we imitate imperfect experimental
calibrations, so that the actual parameters of the pulse shapes are
different from the designed ones. We also consider distortion
(``warping'') of the pulse shapes imitating a nonlinear transfer
function between the control pulses and amplitudes $\textbf{t}_{\rm
e(r)}$. Imperfections due to Gaussian filtering of the pulse shapes,
additional noise, and dissipative losses will also be discussed.

We analyze the effect of imperfections using numerical integration
of the evolution equations (\ref{lg})--(\ref{eq-A}). As the ideally
designed procedure we choose Eqs.\ (\ref{tem})--(\ref{trec-2}) with
$|\textbf{t}_{\rm e,max}|=|\textbf{t}_{\rm r,max}|=0.05$, assuming
the quarter-wavelength resonators with frequency $\omega_{\rm
e}/2\pi=\omega_{\rm r}/2\pi=6$ GHz, so that the round-trip time is
$\tau_{\rm rt,e}=\tau_{\rm rt,r}=\pi/\omega_{\rm e(r)}=1/12$ ns and
the buildup/leakage time is $\tau_{\rm e}=\tau_{\rm r}=\tau =33.3$
ns. The duration of the procedure $t_{\rm f}$ is chosen from Eq.\
(\ref{ineff-2}), using two design values of the efficiency:
$\eta_{\rm d}=0.99$ and $\eta_{\rm d}=0.999$; the corresponding
durations are $t_{\rm f}=307.0$ ns and 460.5 ns. The time $t_{\rm
m}$ is in the middle of the procedure: $t_{\rm m}=t_{\rm f}/2$. In
the simulations we use $G(0)=1$, $B(0)=0$, and calculate the
efficiency as $\eta=|B(t_{\rm f})/G(0)|^2$. Note that the values of
$|\textbf{t}_{\rm e(r),max}|$ and $\omega_{\rm e(r)}$ affect the
duration of the procedure, but do not affect the results for the
efficiency presented in this section (except for the filtering
effect).

\subsection{Variation of maximum transmission amplitudes
$\textbf{t}_{\rm e, max}$ and $\textbf{t}_{\rm r, max}$}

Let us assume that the transmission amplitudes are still described
by the pulse shapes \eqref{tem}--\eqref{trec-2}, but with slightly
different parameters,
    \begin{align}
& \textbf{t}_{\rm e}^{\rm a}(t)=\frac{\textbf{t}_{\rm e, max}^{\rm
a}\sqrt{\tau_{\rm e}^{\rm a}/\tau_{\rm r}^{\rm a}}}{\sqrt{(
1+\tau_{\rm e}^{\rm a}/\tau_{\rm r}^{\rm a}) \exp [(t_{\rm m}^{\rm
a,e}-t)/\tau_{\rm r}^{\rm a}]-1}} , \,\,\, t\leq t_{\rm m}^{\rm
a,e},
    \label{t-e-a}\\
& \textbf{t}_{\rm r}^{\rm a}(t)=\frac{\textbf{t}_{\rm r, max}^{\rm
a}\sqrt{\tau_{\rm r}^{\rm a}/\tau_{\rm e}^{\rm a}}}{\sqrt{(
1+\tau_{\rm r}^{\rm a}/\tau_{\rm e}^{\rm a}) \exp [(t-t_{\rm m}^{\rm
a,r})/\tau_{\rm e}^{\rm a}]-1}}, \,\,\, t\geq t_{\rm m}^{\rm a,r} ,
    \label{t-r-a}\end{align}
so that the ``actual'' parameters $\textbf{t}_{\rm e, max}^{\rm a}$,
$\textbf{t}_{\rm r, max}^{\rm a}$, $\tau_{\rm e}^{\rm a}$,
$\tau_{\rm r}^{\rm a}$, $t_{\rm m}^{\rm a,e}$, and $t_{\rm m}^{\rm
a,r}$ are somewhat different from their design values
$\textbf{t}_{\rm e, max}$, $\textbf{t}_{\rm r, max}$, $\tau_{\rm
e}$, $\tau_{\rm r}$, and $t_{\rm m}$. The transmission amplitudes
are kept at their maxima $\textbf{t}_{\rm e, max}^{\rm a}$ and
$\textbf{t}_{\rm r, max}^{\rm a}$ after/before the possibly
different mid-times $t_{\rm m}^{\rm a,e}$ and $t_{\rm m}^{\rm a,r}$.
We will analyze the effect of inaccurate parameters one by one.

\begin{figure}[t]
\includegraphics[width=8cm]{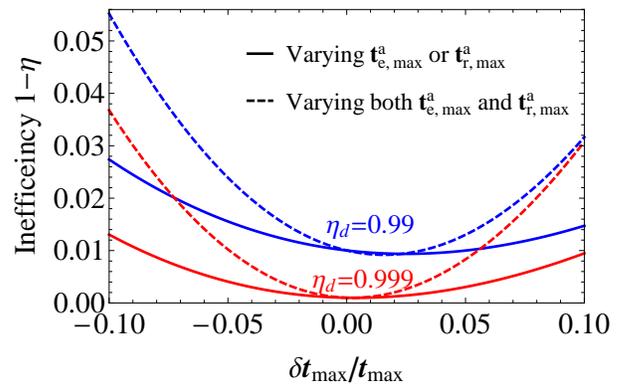}
\caption{Inefficiency $1-\eta$ of the state transfer procedure as a
function of relative variation of the maximum transmission
amplitudes $\delta \textbf{t}_{\rm max}/\textbf{t}_{\rm max} =
(\textbf{t}_{\rm e(r), max}^{\rm a}-\textbf{t}_{\rm e(r),
max})/\textbf{t}_{\rm e(r), max}$ for design efficiencies $\eta_{\rm
d}=0.99$ (blue curves) and $0.999$ (red curves). The maximum
transmission amplitudes $\textbf{t}_{\rm e, max}$ and
$\textbf{t}_{\rm r, max}$ are either varied simultaneously (dashed
curves) or one of them is kept at the design value (solid curves).
The superscript ``a'' indicates an ``actual'' value, different from the
design value.
   } \label{tmax}
\end{figure}

    First, we assume that only the maximum amplitudes
are inaccurate, $\textbf{t}_{\rm e, max}^{\rm a}=\textbf{t}_{\rm e,
max}+\delta \textbf{t}_{\rm e,max}$ and $\textbf{t}_{\rm r,
max}^{\rm a}=\textbf{t}_{\rm r, max}+\delta \textbf{t}_{\rm r,max}$,
while other parameters are equal to their design values. (We change
only the absolute values of $\textbf{t}_{\rm e, max}$ and
$\textbf{t}_{\rm r, max}$, because their phases affect only the
correctable final phase $\varphi_{\rm f}$ but do not affect the
efficiency $\eta$.) In Fig.\ \ref{tmax} we show the numerically
calculated inefficiency $1-\eta$ of the state transfer as a function
of the variation in maximum transmission amplitude $\delta
\textbf{t}_{\rm max}/\textbf{t}_{\rm max}$, with the solid lines
corresponding to variation of only one maximum amplitude, $\delta
\textbf{t}_{\rm e, max}/\textbf{t}_{\rm e, max}$ or $\delta
\textbf{t}_{\rm r, max}/\textbf{t}_{\rm r, max}$ (the results are
the same), and the dashed lines corresponding to variation of both
of them, $\delta \textbf{t}_{\rm e, max}/\textbf{t}_{\rm e,
max}=\delta \textbf{t}_{\rm r, max}/\textbf{t}_{\rm r, max}$. The
blue (upper) lines are for the case of design efficiency $\eta_{\rm
d}=0.99$ and the red (lower) lines are for $\eta_{\rm d}=0.999$.

We see that deviations of the actual maximum amplitudes
$\textbf{t}_{\rm e, max}^{\rm a}$ and $\textbf{t}_{\rm r, max}^{\rm
a}$ from their design values $\textbf{t}_{\rm e, max}$ and
$\textbf{t}_{\rm r, max}$ increase the inefficiency of the state
transfer [essentially because of the inconsistency between
$\textbf{t}_{\rm e(r), max}^{\rm a}$ and $\tau_{\rm e(r)}^{\rm a}$].
However, the effect is not very significant, with the additional
inefficiency of less than 0.006 when one of the parameters deviates
by $\pm 5\%$ and less than 0.02 when both of them deviate by $\pm
5\%$. The curves in Fig.\  \ref{tmax} are approximately parabolic,
with a growing asymmetry for larger $1-\eta_{\rm d}$.

For the case $\eta_{\rm d}\approx 1$ the numerical results for the
additional inefficiency $-\delta \eta=\eta_{\rm d}-\eta$ can be
approximately fitted by the formula
\begin{align}\label{bothtm}
  -\delta\eta \approx \left( \frac{\delta \textbf{t}_{\rm e, max}}{\textbf{t}_{\rm e, max}}\right)^2  \!+ \left( \frac{\delta \textbf{t}_{\rm r, max}}{\textbf{t}_{\rm r, max}}\right)^2 \!
  +1.25\frac{\delta \textbf{t}_{\rm e, max}}{\textbf{t}_{\rm e, max}}\frac{\delta \textbf{t}_{\rm r, max}}{\textbf{t}_{\rm r,max}},
  \end{align}
which we obtained by changing the maximum amplitudes symmetrically,
antisymmetrically, and separately. Note that in the ideal procedure
we assumed $|\textbf{t}_{\rm e, max}|=|\textbf{t}_{\rm r, max}|$.

The main result here is that the state transfer is quite robust
against the small variation of the transmission amplitudes. We expect that experimentally these parameters can be calibrated with accuracy of a few per cent or better; the related inefficiency of the transfer protocol is very small.

\subsection{Variation of buildup/leakage times $\tau_{\rm e}$ and $\tau_{\rm r}$}

\begin{figure}[t]
\includegraphics[width=8cm]{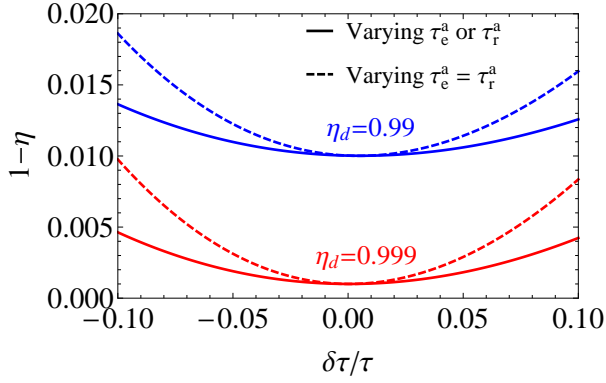}
 \caption{Dependence of the inefficiency $1-\eta$ on relative
variation of the buildup/leakage time $\delta\tau_{\rm e(r)}/\tau
=(\tau_{\rm e(r)}^{\rm a}-\tau)/\tau$ for design efficiencies
$\eta_{\rm d}=0.99$ (blue curves) and $0.999$ (red curves), assuming
$\tau_{\rm e}=\tau_{\rm r}=\tau$. The buildup/leakage times
$\tau_{\rm e}^{\rm a}$ and $\tau_{\rm r}^{\rm a}$ are varied either
simultaneously (dashed curves) or one of them is kept at the design
value (solid curves).
   } \label{fig-tau}
\end{figure}

Now let us assume that in Eqs.\ (\ref{t-e-a}) and (\ref{t-r-a}) only
the buildup/leakage time parameters are slightly inaccurate,
$\tau_{\rm e}^{\rm a}=\tau +\delta \tau_{\rm e}$ and $\tau_{\rm
r}^{\rm a}=\tau+\delta \tau_{\rm r}$ (we assume that in the ideal
procedure $\tau_{\rm e}=\tau_{\rm r}=\tau$), while other parameters
are equal to their design values. The transfer inefficiency as a
function of the relative deviations $\delta \tau_{\rm e(r)}/\tau$ is
shown in Fig.\ \ref{fig-tau} for the design efficiencies $\eta_{\rm
d}=0.99$ (blue lines) and $0.999$ (red lines). For the solid lines
only one of the buildup/leakage times is varied (the results
coincide), while for the dashed lines both parameters are varied
together, $\delta \tau_{\rm e}=\delta \tau_{\rm r}$. As we see, $\pm
5\%$ variation of one of the buidup/leakage times increases the
inefficiency by less than 0.001, and by less than 0.0025 if the both
times are varied by $\pm 5\%$.

The approximately parabolic dependences shown in Fig.\ \ref{fig-tau} can be numerically fitted by the formula for the additional inefficiency  $-\delta\eta$,
    \begin{align}\label{tauboth}
 -\delta\eta\approx 0.34\left[ \left(\frac{\delta\tau_{\rm e}}{\tau}
\right)^2+\left(\frac{\delta\tau_{\rm r}}{\tau}\right)^2\right]+
0.12 \frac{\delta \tau_{\rm e}}{\tau}\frac{\delta \tau_{\rm
r}}{\tau},
     \end{align}
which was again obtained by varying $\delta\tau_{\rm e}$ and $\delta\tau_{\rm r}$ symmetrically, antisymmetrically, and separately.
Most importantly, we see that the transfer procedure is robust against small deviations of the buildup/leakage times. (In an experiment we expect not more than a few per cent inaccuracy for these parameters.)

\subsection{Variation of mid-times $t_{\rm m}^{\rm a,e}$ and $t_{\rm m}^{\rm a,r}$}

Ideally, the pulse shapes $\textbf{t}_{\rm e}(t)$ and
$\textbf{t}_{\rm r}(t)$ should switch from increasing/decreasing
parts to constants at the same time $t_{\rm m}$, exactly in the
middle of the procedure. However, due to imperfectly calibrated
delays in the lines delivering the signals to the couplers, this
change may occur at slightly different actual times  $t_{\rm m}^{\rm
a,e}$ and $t_{\rm m}^{\rm a,r}$, which are also not necessarily
exactly in the middle of the procedure. Let us assume that
$\textbf{t}_{\rm e}(t)$ and $\textbf{t}_{\rm r}(t)$ are given by
Eqs.\ (\ref{t-e-a}) and (\ref{t-r-a}) with slightly inaccurate times
$t_{\rm m}^{\rm a,e}$ and $t_{\rm m}^{\rm a,r}$, while other
parameters are equal to their design values.

\begin{figure}[t]
\includegraphics[width=8cm]{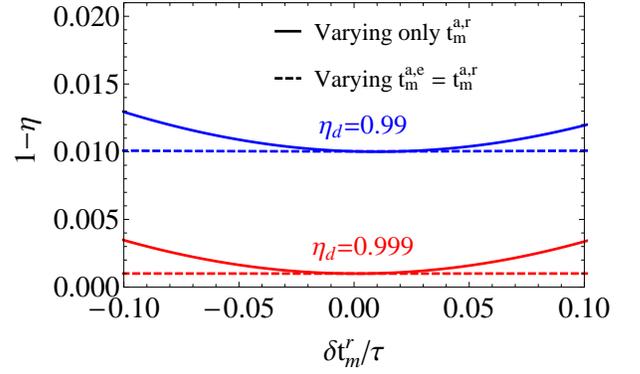}
\caption{Inefficiency $1-\eta$ as a function of the mid-time
shift $\delta t_{\rm m}^{\rm r}=t_{\rm m}^{\rm a, r}-t_{\rm m}$
normalized by the buildup/leakage time $\tau$. The mid-time $t_{\rm
m}^{\rm a,e}$ is either varied equally (dashed curves) or kept
constant (solid curves). The results for varying only $t_{\rm
m}^{\rm a, e}$ are the same as the solid curves up to the sign
change, $\delta t_{\rm m}^{\rm r} \leftrightarrow -\delta t_{\rm
m}^{\rm e}$.
      } \label{figtm}
\end{figure}

Solid lines in Fig.\ \ref{figtm} show the dependence of the transfer
inefficiency $1-\eta$ on the shift of the mid-time $\delta t_{\rm
m}^{\rm r}=t_{\rm m}^{\rm a,r}-t_{\rm m}$, which is normalized by
the buildup/leakage time $\tau$. Blue and red lines are for the
design efficiencies $\eta_{\rm d}=0.99$ and 0.999, respectively. The
case when only $t_{\rm m}^{\rm a,e}$ is changed is similar to what
is shown by the solid lines up to the mirror symmetry, $\delta
t_{\rm m}^{\rm r} \leftrightarrow -\delta t_{\rm m}^{\rm e}$. The
dashed lines show the case when both mid-times are shifted
simultaneously, $t_{\rm m}^{\rm a,e}=t_{\rm m}^{\rm a,r}$.

We see that when $t_{\rm m}^{\rm a,e}$ and $t_{\rm m}^{\rm a,r}$
coincide, there is practically no effect of the shift. This is
because in this case the change is only due to slightly unequal
durations $t_{\rm m}^{\rm a}$ and $t_{\rm f}-t_{\rm m}^{\rm a}$. A
non-zero time mismatch $t_{\rm m}^{\rm a,e}-t_{\rm m}^{\rm a,r}$ has
a much more serious effect because the reflection cancellation
(\ref{F-form1}) becomes significantly degraded in the middle of
the procedure, where the propagating field is at its maximum.

The numerical fit to a quadratic dependence gives
\begin{equation}\label{tm}
  -\delta\eta\approx 0.25 \left(\frac{\delta t_{\rm m}^{\rm a,e}-\delta t_{\rm m}^{\rm a,r}}{\tau}\right)^2.
\end{equation}
For $\tau =33.3$ ns this means that $\sim$3 ns time mismatch leads to only $2\times10^{-3}$ increase in inefficiency. Such robustness to the time mismatch is rather surprising. It can be qualitatively explained in the following way. The relative imperfection of the back-reflection cancellation (\ref{F-form1}) is approximately $(\delta t_{\rm m}^{\rm a,e}-\delta t_{\rm m}^{\rm a,r})/\tau$ in the middle of the procedure; however, the lost energy of the back-reflected field scales quadratically. Therefore, we can explain Eq.\ (\ref{tm}) up to a numerical factor. In an experiment we expect that the time mismatch can be made smaller than 1 ns; the corresponding inefficiency is almost negligible.

\subsection{Pulse-shape warping}

As another possible imperfection of the ideal time-dependences
$\textbf{t}_{\rm e}(t)$ and $\textbf{t}_{\rm r}(t)$, we consider a
nonlinear deformation (``warping'') with the form
\begin{equation}\label{tnw}
  \textbf{t}_{ j}^{\rm a}(t)=\textbf{t}_{j}(t) \left[ 1+\alpha_{j}
  \frac{\textbf{t}_{j}(t)-\textbf{t}_{j,\rm max}}{\textbf{t}_{j,\rm max}}
  \right],\,\,\,j=\text{e,\,r},
\end{equation}
where $\alpha_{\rm e}$ and $\alpha_{\rm r}$ are the warping
parameters, which determine the strength of the deformations.  Note
that this deformation does not affect maximum values
$\textbf{t}_{\rm e(r), max}$ and the values close to zero; it
affects only intermediate values. The deformation imitates nonlinear
(imperfectly compensated) conversion from experimental control
signals into transmission amplitudes.

\begin{figure}[t]
\includegraphics[width=8cm]{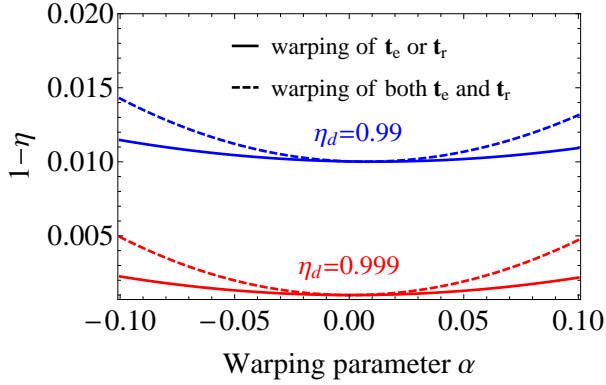}
  \caption{Dependence of the inefficiency $1-\eta$ on the warping
parameters $\alpha_{\rm e}$ and $\alpha_{\rm r}$, introduced in Eq.\
\eqref{tnw} to describe the pulse shape distortion, for design
efficiencies $\eta_{\rm d}=0.99$ (blue curves) and $0.999$ (red
curves). The solid curves show the case when only one warping
parameter is non-zero (the results coincide); the dashed curves are
for the case $\alpha_{\rm e}=\alpha_{\rm r}$.
   } \label{warp}
\end{figure}

The inefficiency increase due to the warping of the transmission
amplitude pulse shapes is illustrated in Fig.~\ref{warp}. Solid
lines show the case when only $\alpha_{\rm e}$ or $\alpha_{\rm r}$
is non-zero (the results coincide), while the dashed lines show the
case $\alpha_{\rm e}=\alpha_{\rm r}$. We see that for $\alpha_{\rm
e}=\alpha_{\rm r}=0.05$ the inefficiency increases by $\sim 10^{-3}$
for both design efficiencies $\eta_{\rm d}=0.99$ and $0.999$.
Similar to the variation of other parameters, the inefficiency due
to the warping effect has a quadratic dependence on the warping
parameters $\alpha_{\rm e}$ and $\alpha_{\rm r}$. The numerical
fitting for small $|\alpha_{\rm e(r)}|$ and $\eta\approx 1$ gives
      \begin{equation}
-\delta\eta\approx 0.22(\alpha_{\rm e}^2+\alpha_{\rm r}^2)+ 0.12\alpha_{\rm e}\alpha_{\rm r}.
     \label{d-eta-warp}\end{equation}

Again, this result shows that the state transfer is robust to
distortion of the couplers' transmission amplitude pulse shapes. We do not expect that uncompensated experimental nonlinearities will follow Eq.\ (\ref{tnw}) exactly, since this equation only imitates a nonlinear conversion. However, very crudely, we would expect that $|\alpha_{\rm e(r)}|<0.05$ is a realistic  experimental estimate.

\subsection{Smoothing by a Gaussian filter}\label{gau_smooth}

In an actual experiment the designed pulse shapes for the
transmission amplitudes of the tunable couplers given by Eqs.\
\eqref{tem}--\eqref{trec-2} will pass through a filter. Here we
convolve the transmission amplitudes with a Gaussian function to
simulate the experimental filtering, so the actual transmission
amplitudes are
\begin{equation}\label{smooth}
  \textbf{t}_{j}^{\rm a}(t)=\frac{1}{\sqrt{2\pi}\, \sigma}
  \int_{-\infty}^{\infty} e^{-(t-t')^2/2\sigma^2}\textbf{t}_{j}
  (t') \, dt', \,\,\, j=\text{e,\,r}, \quad
\end{equation}
where $\sigma$ is the time-width of the Gaussian filter. The
filtering smooths out the kinks at the middle of the procedure and
slightly lowers the initial and final values of $\textbf{t}_{\rm e}$
and $\textbf{t}_{\rm r}$. The change in transmission amplitudes
translates into a decrease in the state transfer efficiency. Note
that the smoothing reduces the energy loss at the beginning and end
of the procedure, but causes an increased energy loss at the middle
of the procedure, thus  increasing the procedure inefficiency
overall.

\begin{figure}[t]
\includegraphics[width=8cm]{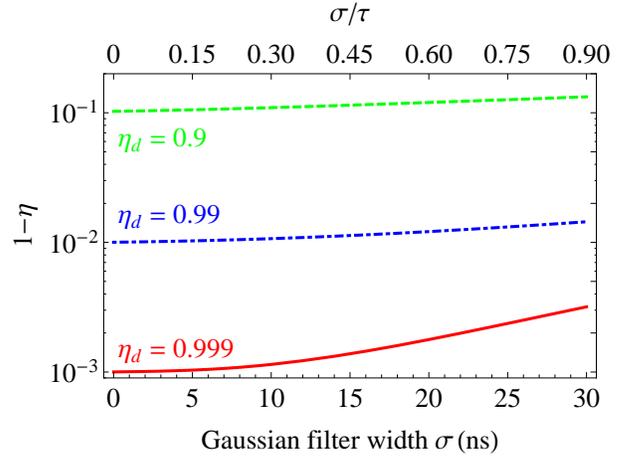}
\caption{Inefficiency $1-\eta$ as a function of the width of a
Gaussian filter $\sigma$ (in ns) for design efficiencies $\eta_{\rm
d}=0.9$ (green dashed curve), $0.99$ (blue dot-dashed curve), and
$0.999$ (red solid curve). We use $\tau = 33.3 \, \rm ns$, as in
Fig.\ \ref{pulse-shapes}.  The upper horizontal axis shows the
normalized value $\sigma/\tau$.
  }
\label{gausmooth}
\end{figure}

The procedure inefficiency with the effect of the Gaussian filtering
of transmission amplitudes is shown in Fig.\ \ref{gausmooth} for the
design efficiencies $\eta_{\rm d}=0.9$, $0.99$, and $0.999$. Rather
surprisingly, the effect is very small, so that filtering with
$\sigma =10$ ns does not produce a noticeable increase of the
inefficiency, and even with $\sigma = 30$ ns (which is close to the
buildup/leakage time) the effect is still small. Such robustness to
the filtering can be qualitatively understood in the same way as the
robustness to the mismatch between the mid-times $\textbf{t}_{\rm
e}(t)$ and $\textbf{t}_{\rm r}(t)$ discussed above. Note that
experimentally \cite{Hofheinz-09} $\sigma$ is on the order of
1 ns, so the effect of the filter on the efficiency should be
negligible.

\subsection{Noisy transmission amplitudes}

In experiment the pulse shapes $\textbf{t}_{\rm e}(t)$ and
$\textbf{t}_{\rm r}(t)$ may contain noise. We model this noise by
replacing the designed pulse shapes $\textbf{t}_{\rm e}(t)$ and
$\textbf{t}_{\rm r}(t)$ with ``actual'' shapes as
\begin{equation}\label{noise1}
  \textbf{t}_{j}^{\rm a}(t)=\textbf{t}_{j}(t) [1+a\, \xi_j(t)], \,\,\,
  j=\text{e, r},
\end{equation}
where $a$ corresponds to the dimensionless noise amplitude and
$\xi_{\rm e}(t)$ and $\xi_{\rm r}(t)$ are mutually uncorrelated
random processes. We generate each $\xi (t)$ numerically in the
following way. First, we choose a time step $dt$ and generate $\xi
(t)$ at discrete time moments $t=n\, dt$ (with integer $n$) as
Gaussian-distributed random numbers with zero mean and unit standard
deviation. After that we create a smooth function $\xi (t)$ passing
through these points by polynomial interpolation. Since the noise
contribution in Eq.\ (\ref{noise1}) scales with the transmission
amplitude $\textbf{t}_{j}$, we call it a multiplicative
noise. Besides that, we also use a model of an additive noise
defined as
\begin{equation}\label{noise2}
  \textbf{t}_{j}^{\rm a}(t)=\textbf{t}_{j}(t)+
  a \, \textbf{t}_{j,\rm max}\, \xi_j(t), \,\,\,
  j=\text{e,\,r},
\end{equation}
where the relative amplitude $a$ is now compared with the maximum
value $\textbf{t}_{j,\rm max}$, while each $\xi(t)$ is generated in
the same way. Note that for sufficiently small $dt$ the noise
$\xi(t)$ is practically white at low frequency; its variance
$\overline{\xi^2}$ does not depend on $dt$, and therefore the low
frequency spectral density is proportional to $dt$ (the effective
cutoff frequency scales as $dt^{-1}$). Also note that the variance
$\overline{\xi^2}$ somewhat depends on the method of interpolation
used to generate $\xi(t)$. For the default interpolation method in
Mathematica, which we used (polynomial interpolation of order three), $\overline{\xi^2}\approx 0.78$.

\begin{figure}[t]
\includegraphics[width=8cm]{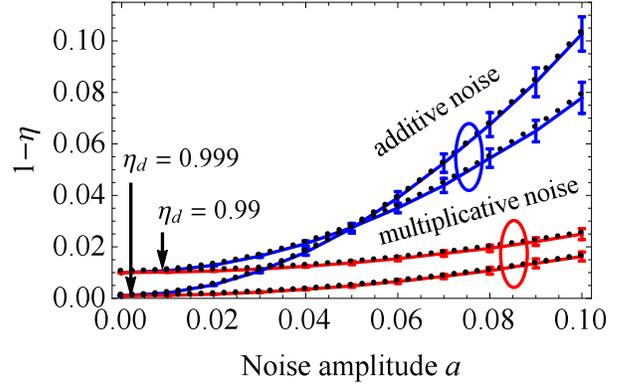}
\caption{Solid lines: inefficiency $1-\eta$ averaged over 100 random
noise realizations, as a function of the dimensionless noise
amplitude $a$, for the multiplicative noise (red lines, bottom),
Eq.\ (\ref{noise1}), and the additive noise (blue lines, top), Eq.\
(\ref{noise2}); both with $\overline{\xi^2}=0.78$. The error bars
show the standard deviation for some values of $a$. The results are
shown for $\eta_{\rm d}=0.99$ and $0.999$. In the simulation we used
the time step $dt=1\, \rm ns$ and parameters of the procedure in
Fig.\ \ref{pulse-shapes} ($\tau = 33.3\, \rm ns$). Black dotted
lines (practically coinciding with the solid lines) are calculated
by replacing the noise with the effective increase of the leakage
rates $\kappa_{\rm e(r)}$ (see text).
   } \label{noise}
\end{figure}

The numerical results for the transfer inefficiency $1-\eta$ in the
presence of noise are shown in Fig.\ \ref{noise} as a function of
the dimensionless amplitude $a$. We used the time step $dt=
1~\text{ns}$ and design efficiencies $\eta_{\rm d}=0.99$ and
$\eta_{\rm d}=0.999$. The results are averaged over 100 random
realizations; we show the average values by the solid lines and also
show the standard deviations at some values of $a$. Red lines
correspond to the multiplicative noise, while blue lines correspond
to the additive noise. As expected, the additive noise leads to larger
inefficiency than the multiplicative noise with the same amplitude,
because of larger noise at the non-constant part of the pulse shape.

It is somewhat surprising that, as we checked numerically, the
average results shown in Fig.\ \ref{noise} by the solid lines
practically do not depend on the choice of the time step $dt$, as
long as $dt\ll \tau_{\rm e(r)}$ (even though in our simulations $dt$ affects the noise spectral density). The error bars, however, scale with
$dt$ as $\sqrt{dt}$. This behavior can be understood in the
following way. In the evolution equations (\ref{lg})--(\ref{eq-A}),
the noise in $\textbf{t}_{\rm e}(t)$ and $\textbf{t}_{\rm r}(t)$
affects the leakage rates $\kappa_{\rm e}\propto |\textbf{t}_{\rm
e}|^2$ and $\kappa_{\rm r}\propto |\textbf{t}_{\rm r}|^2$ of the two
resonators, and also affects the transfer term $\sqrt{\kappa_{\rm
e}\kappa_{\rm r}}\, A \propto |\textbf{t}_{\rm e}\textbf{t}_{\rm
r}|$. On average the transfer term does not change (because the
noises of $\textbf{t}_{\rm e}(t)$ and $\textbf{t}_{\rm r}(t)$ are
uncorrelated); however, the average values of $|\textbf{t}_{\rm
e}|^2$ and $|\textbf{t}_{\rm r}|^2$ change as $\langle
|\textbf{t}_{\rm e(r)}^{\rm a}|^2\rangle =|\textbf{t}_{\rm e(r)}|^2
(1+a^2\, \overline{\xi^2})$ for the model of Eq.\ (\ref{noise1}) and
as $\langle |\textbf{t}_{\rm e(r)}^{\rm a}|^2\rangle
=|\textbf{t}_{\rm e(r)}|^2 + a^2 \, |\textbf{t}_{\rm e(r),max}|^2\,
\overline{\xi^2}$ for the model of Eq.\ (\ref{noise2}). Therefore,
on average we expect dependence on $a^2\, \overline{\xi^2}$ (a
second-order effect), but no dependence on $dt$, as long as it is
sufficiently small. In contrast, the error bars in Fig.\ \ref{noise}
should depend on $dt$ because the transfer term $\sqrt{\kappa_{\rm
e}\kappa_{\rm r}}\, A$ fluctuates linearly in $\xi$. Since the
low-frequency spectral density of $\xi(t)$ scales as $dt$, the
typical fluctuation should scale as $\sqrt{dt}$, thus explaining
such dependence for the error bars in Fig.\ \ref{noise}.
Simply speaking, for a wide-bandwith noise the average value of $1-\eta$ depends on the overall r.m.s.\ value of the noise, while the fluctuations of $1-\eta$ (from run to run) depend on the spectral density of the noise at relatively low frequencies ($\alt \tau^{-1}$).
 Note that the noise can increase or decrease the inefficiency compared to its
average value; however, it always increases the inefficiency in
comparison with the case without noise (as we see from Fig.\
\ref{noise}, even if we increase $dt$ from 1 ns to about the
buildup/leakage time of 33.3 ns, the error bars, increased by the
factor $\sqrt{33.3}$, are still significantly less than the increase
of inefficiency compared with the design value).

We have checked this explanation of the noise effect on the average
inefficiency by replacing the fluctuating evolution equations
(\ref{lg})--(\ref{eq-A}) with non-fluctuating equations, in which
the transfer term $\sqrt{\kappa_{\rm e}\kappa_{\rm r}}\, A$ does not
change, while the leakage rates $\kappa_{\rm e}$ and $\kappa_{\rm
r}$ are multiplied either by $1+a^2\, \overline{\xi^2}$ (for
multiplicative noise) or by $1+a^2\,
\overline{\xi^2}(\textbf{t}_{\rm e(r),max}/\textbf{t}_{\rm e(r)})^2$
for the additive noise. The results are shown in Fig.\ \ref{noise} by
the dotted lines; we see that they almost coincide with the solid
lines, thus confirming our explanation. We have also used several
interpolation methods, which give somewhat different
$\overline{\xi^2}$, and checked that the direct simulation with
fluctuations and use of the non-fluctuating equations still give the
same results.

As can be seen from Fig.\ \ref{noise}, the average inefficiency
depends approximately quadratically on the noise amplitude $a$ for
both additive and multiplicative noise. The additional inefficiency
$-\delta\eta$ can be fitted numerically as
\begin{equation}\label{noise-fit}
  -\delta\eta\approx c_{\rm n} a^2 \, \overline{\xi^2},
\end{equation}
where $c_{\rm n}\approx 2$ for the multiplicative noise and $c_{\rm
n}\approx 2\, \ln \frac{1}{1-\eta_{\rm d}}$ for the additive noise.
Note that for the additive noise $c_{\rm n}$ increases with decreasing
design inefficiency $1-\eta_{\rm d}$, so the blue lines in Fig.\
\ref{noise} intersect. This is because a smaller $1-\eta_{\rm d}$
requires a longer procedure duration $t_{\rm f}$, causing more loss
due to additional leakage of the resonators caused by fluctuating
$\textbf{t}_{\rm e(r)}$.

The value of $c_{\rm n}$ for the additive noise can be derived
analytically in the following way. As discussed above, the noise
essentially increases the resonator leakages, $\kappa^{\rm a}_{\rm
e(r)}(t)=\kappa_{\rm e(r)}(t)+a^2\, \overline {\xi^2}/\tau$, without
increasing the transferred field; therefore, it is equivalent to the
effect of energy relaxation with $T_1=\tau/(a^2\,
\overline{\xi^2})$. Consequently (see below), the efficiency
decreases as $\eta=\eta_{\rm d} \exp (-t_{\rm f}/T_1)=\eta_{\rm d}
\exp (-2a^2\, \overline{\xi^2}\ln\frac{1}{1-\eta_{\rm d}})$ [see
Eq.\ (\ref{ineff-1}) for $t_{\rm f}$], and the linear expansion of
the exponent in this formula reproduces Eq.\ (\ref{noise-fit}) with
$c_{\rm n}= 2\, \ln \frac{1}{1-\eta_{\rm d}}$.

The value of $c_{\rm n}$ for the multiplicative noise can be
derived in a somewhat similar way. Now $\kappa^{\rm a}_{\rm
e}(t)=\kappa_{\rm e}(t) (1+a^2\, \overline {\xi^2})$, so the
additional leakage of the emitting resonator consumes the fraction
$a^2\, \overline {\xi^2}$ of the transmitted energy. Using the
time-reversal picture, we see that an analogous increase of the
receiving resonator leakage, $\kappa^{\rm a}_{\rm r}(t)=\kappa_{\rm
r}(t) (1+a^2\, \overline {\xi^2})$, emits (back-reflects) into the
transmission line the fraction $a^2\, \overline {\xi^2}$ of the
final energy $|B(t_{\rm f})|^2$. Combining these two losses, we
obtain $\eta=\eta_{\rm d}(1-2a^2\,\overline{\xi^2})$, which for
$\eta_{\rm d}\approx 1$ reproduces Eq.\ (\ref{noise-fit}) with
$c_{\rm n}= 2$.

Overall, the efficiency decrease due to the multiplicative noise is
not strong; for example, to keep $-\delta\eta < 0.01$ we need the
relative r.m.s.\ fluctuations of  $\textbf{t}_{\rm e(r)}$ to be less
than 7\%. The (additive) fixed-amplitude fluctuations of $\textbf{t}_{\rm
e(r)}$ can be more problematic, because the inability to keep
$\textbf{t}_{\rm e(r)}$ near zero at the initial or final stage of
the procedure leads to loss during most of the (relatively long)
procedure. For example, for $\eta_{\rm d}=0.99$ and $-\delta\eta <
0.01$, we need the r.m.s.\ fluctuations of  $\textbf{t}_{\rm e(r)}$
to be less than 3\% of $\textbf{t}_{\rm e(r), max}$.

\subsection{Effect of dissipation}\label{sec-diss}

For completeness let us discuss here the effect of dissipation by
assuming imperfect transfer through the transmission line,
$\eta_{\rm tl}\neq 1$, and finite energy relaxation times
$T_{1,\rm e}$ and $T_{1,\rm r}$ in the evolution equations
(\ref{lg})--(\ref{eq-A}), while the pulse shapes $\textbf{t}_{\rm
e}(t)$ and $\textbf{t}_{\rm r}(t)$ are assumed to be ideal.

The effect of imperfect $\eta_{\rm tl}$ is easy to analyze, since
the transmitted (classical) field is simply multiplied by
$\sqrt{\eta_{\rm tl}}$. Therefore, the transfer procedure efficiency
is simply multiplied by $\eta_{\rm tl}$, so that $\eta = \eta_{\rm
tl}\eta_{\rm d}$. (Recall that we neglect multiple reflections.)

The effect of energy relaxation in the resonators is also very simple if
$T_{1,\rm r}=T_{1,\rm e}=T_1$. Then the (classical) field decays
equally everywhere, and therefore, after the procedure duration
$t_{\rm f}$, the energy acquires the factor $\exp(-t_{\rm f}/T_1)$,
so that $\eta=\eta_{\rm d}\exp(-t_{\rm f}/T_1)$. The analysis of the
case when $T_{1,\rm r}\neq T_{1,\rm e}$ is not so obvious. We have
analyzed this case numerically and found that the two resonators
bring the factors $\exp(-t_{\rm f}/2T_{1,\rm e})$ and $\exp(-t_{\rm
f}/2T_{1,\rm r})$, respectively.

Combining the effects of dissipation in the resonators and
transmission line, we obtain
    \be
    \eta = \eta_{\rm d} \eta_{\rm tl} \exp(-t_{\rm f}/2T_{1,\rm e})
    \exp(-t_{\rm f}/2T_{1,\rm r}),
    \label{dissip}\ee
assuming that everything else is ideal.

\section{Multiple reflections}\label{mr}

So far we have not considered multiple reflections of the field that
is back-reflected from the receiving end, by assuming either a very
long transmission line or the presence of a circulator [see Fig.\
\ref{fig-model}(b)]. If there is no circulator and the transmission
line is not very long (as for the state transfer between two on-chip
superconducting resonators), then the back-reflected field bounces
back and forth between the couplers and thus affects the efficiency
of the state transfer. To describe these multiple reflections, we
modify the field equations (\ref{lg})--(\ref{eq-A}) by including the
back-propagating field $F(t)$ into the dynamics, for simplicity
assuming in this section $\Delta\omega_{\rm r}=\Delta\omega_{\rm
e}=0$, $\eta_{\rm tl}=1$, and $T^{-1}_{1,\rm e(r)}=0$:
\begin{eqnarray}
&&  \dot G(t) = -\frac{\kappa_{\rm e}}{2} G(t)+
\frac{\textbf{t}_{\rm e}}{|\textbf{t}_{\rm e}|}
\frac{|\textbf{r}_{\rm e}|}{\textbf{r}_{\rm e}^{\rm in}}
\sqrt{\kappa_{\rm e}} \, e^{i\varphi} F(t-t_{\rm d}),\qquad
     \label{ng} \\
&& \dot B(t) = -\frac{\kappa_{\rm r}}{2} B(t)+\frac{\textbf{t}_{\rm
r}}{|\textbf{t}_{\rm r}|} \sqrt{\kappa_{\rm r}}A(t),\label{nb}
    \\ \label{A-refl}
&&  A(t)=\frac{\textbf{t}_{\rm e}}{|\textbf{t}_{\rm
e}|}\sqrt{\kappa_{\rm e}}G(t) +\frac{\textbf{r}_{\rm e}^{\rm
out}}{|\textbf{r}_{\rm e}|}e^{i\varphi} F(t-t_{\rm d}).
\end{eqnarray}
Here $t_{\rm d}$ is the round-trip delay time ($t_{\rm d}=2l_{\rm
tl}/v$, where $l_{\rm tl}$ is the transmission line length and $v$
is the effective speed of light), $\varphi=\omega_{\rm e(r)} t_{\rm
d}$ is the corresponding phase acquired in the round trip, $F(t)$ is
given by Eq.\ (\ref{F-form2}), $\textbf{r}_{\rm e}^{\rm out}$ is the
reflection amplitude of the emitting resonator coupler from the
transmission line side, and $\textbf{r}_{\rm e}^{\rm in}$ is the
same from the resonator side. Note that we use shifted clocks, so
the propagation is formally infinitely fast in the forward direction
and has velocity $v/2$ in the reverse direction; then the round-trip
delay $t_{\rm d}$ and phase shift $\varphi$ are accumulated in the
back-propagation only; the field $F(t)$ is defined at the receiving
resonator, and it comes to the emitting resonator as $e^{i\varphi}
F(t-t_{\rm d})$. Also note that even though $\varphi$ is
proportional to $t_{\rm d}$, it is better to treat $\varphi$ as an
independent parameter, because the time-delay effects are determined by the ratio
$t_{\rm d}/\tau$, which has a very different scale from $\varphi = (t_{\rm d}/\tau)\, \omega_{\rm e(r)}\tau$, since $\omega_{\rm e(r)}\tau \sim 10^3$.

There is some asymmetry between Eqs.\ (\ref{ng}) and (\ref{nb}) and also between Eqs.\ (\ref{A-refl}) and (\ref{F-form1}), which involves factors $\textbf{r}^{\rm in}_{\rm e(r)}$. This is because in order to keep a simple form of the evolution equations (\ref{lg})--(\ref{eq-A}), we essentially defined $G$ as the field propagating towards the transmission line, while $B$ propagates away from the transmission line.
In this section we still assume that the phases of the transmission
and reflection amplitudes ($\textbf{t}_{\rm e(r)}$ and
$\textbf{r}_{\rm e(r)}^{\rm in (out)}$) do not change with time. For
the tunable couplers of Refs.\ \cite{Yin13,Wen14} (see Appendix B)
the transmission amplitudes $\textbf{t}_{\rm e(r)}$ are mostly
imaginary, the reflection amplitudes $\textbf{r}_{\rm e(r)}^{\rm
in}$ are close to $-1$, and $\textbf{r}_{\rm e(r)}^{\rm out}$ are
somewhat close to $-1$ (recall that $\textbf{t}_{\rm
e}^2/\textbf{r}_{\rm e}^{\rm in}\textbf{r}_{\rm e}^{\rm out}$  and
$\textbf{t}_{\rm r}^2/\textbf{r}_{\rm r}^{\rm in}\textbf{r}_{\rm
r}^{\rm out}$ must be real and negative from unitarity). In simulations
it is easier to redefine the phases of the fields in the resonators
and transmission line, so that $\textbf{t}_{\rm e}$ and
$\textbf{t}_{\rm r}$ are treated as real and positive numbers,
$\textbf{r}_{\rm e}^{\rm in}$ and $\textbf{r}_{\rm r}^{\rm in}$ are
also real and positive (close to $1$), while $\textbf{r}_{\rm e}^{\rm
out}$ and $\textbf{r}_{\rm r}^{\rm out}$ are real and negative (close to
$-1$). In this case Eqs.\ (\ref{F-form2}) and (\ref{A-refl}) become
$F=\sqrt{\kappa_{\rm r}}\, B-A$ and $A(t)=\sqrt{\kappa_{\rm e}}\,
G(t) -e^{-i\varphi} F(t-t_{\rm d})$.

\begin{figure}[t]
\includegraphics[width=8cm]{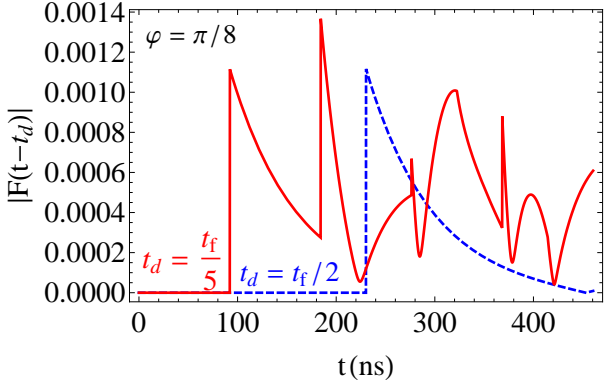}
\caption{Illustration of the back-reflected field $|F(t-t_{\rm d})|$
reaching the emitting resonator at time $t$, for the round-trip
delay time $t_{\rm d}=t_{\rm f}/2$ (blue dashed curve) and $t_{\rm
d}=t_{\rm f}/5$ (red solid curve), assuming the round-trip phase
shift $\varphi=\pi/8$. The kinks represent multiple reflections of
the field emitted at $t=0$. We assumed parameters of Fig.\
\ref{pulse-shapes} ($\eta_{\rm d}=0.999$, $\tau=33.3$ ns, $t_{\rm
f}=460$ ns).
  } \label{f}
\end{figure}

As an example of the dynamics with multiple reflections, in Fig.\
\ref{f} we show the absolute value of the reflected field
$F(t-t_{\rm d})$ (at the emitting resonator) for the procedure shown
in Fig.\ \ref{pulse-shapes} ($\eta_{\rm d}=0.999$, $t_{\rm f}=460$
ns) for the round-trip delays $t_{\rm d}=t_{\rm f}/2$ (blue dashed
curve) and $t_{\rm d}=t_{\rm f}/5$ (red solid curve), assuming
$\varphi=\pi/8$. The kinks represent the successive reflections of
the field emitted at $t=0$. Note that depending on the phase shift
$\varphi$, the resulting contribution of the reflected field into
$B(t_{\rm f})$ can either increase or decrease $|B(t_{\rm f})|^2$,
thus either decreasing or increasing the transfer efficiency $\eta$
(recall that the efficiency $\eta$ is defined disregarding the
resulting phase $\varphi_{\rm f}$, because it can be easily
corrected in an experiment). The effect of multiple reflections
should vanish if $t_{\rm d}\geq t_{\rm f}$, i.e.\ when the
transmission line is sufficiently long.

\begin{figure}[t]
\includegraphics[width=8.5cm]{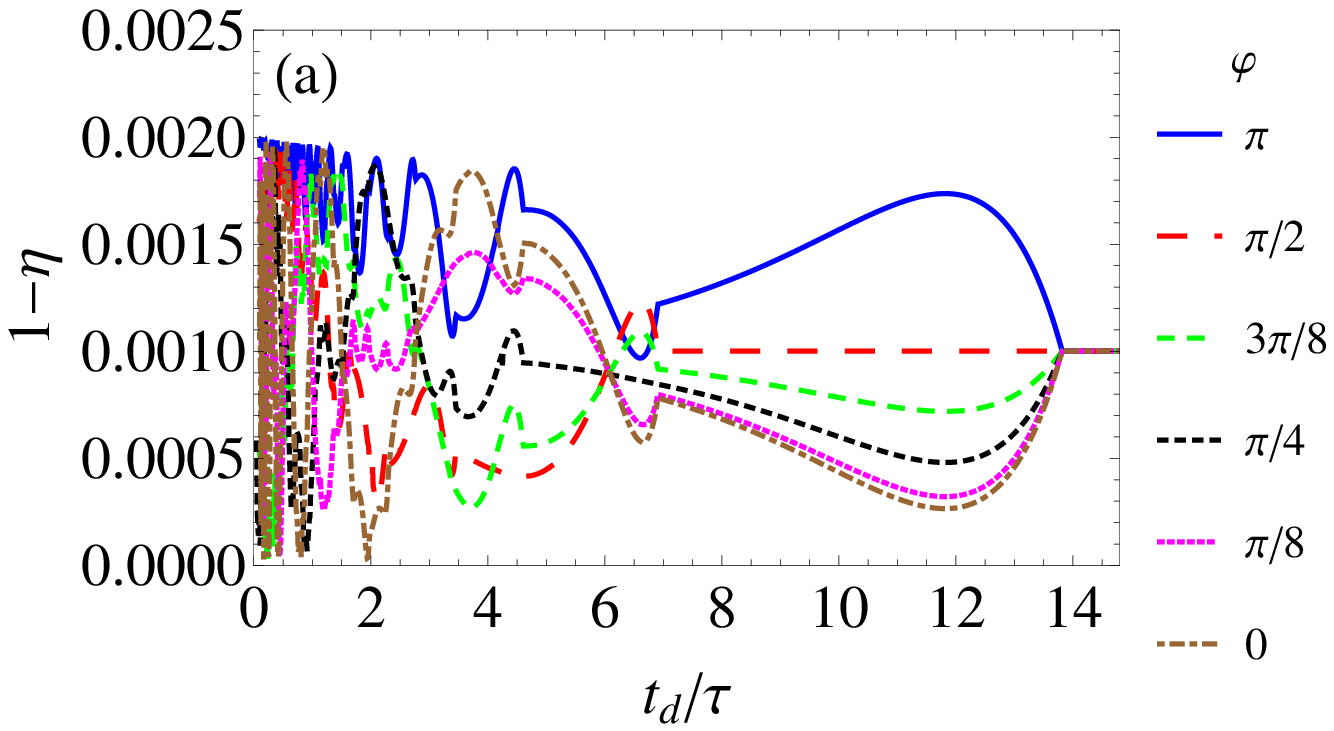}
\includegraphics[width=8.5cm]{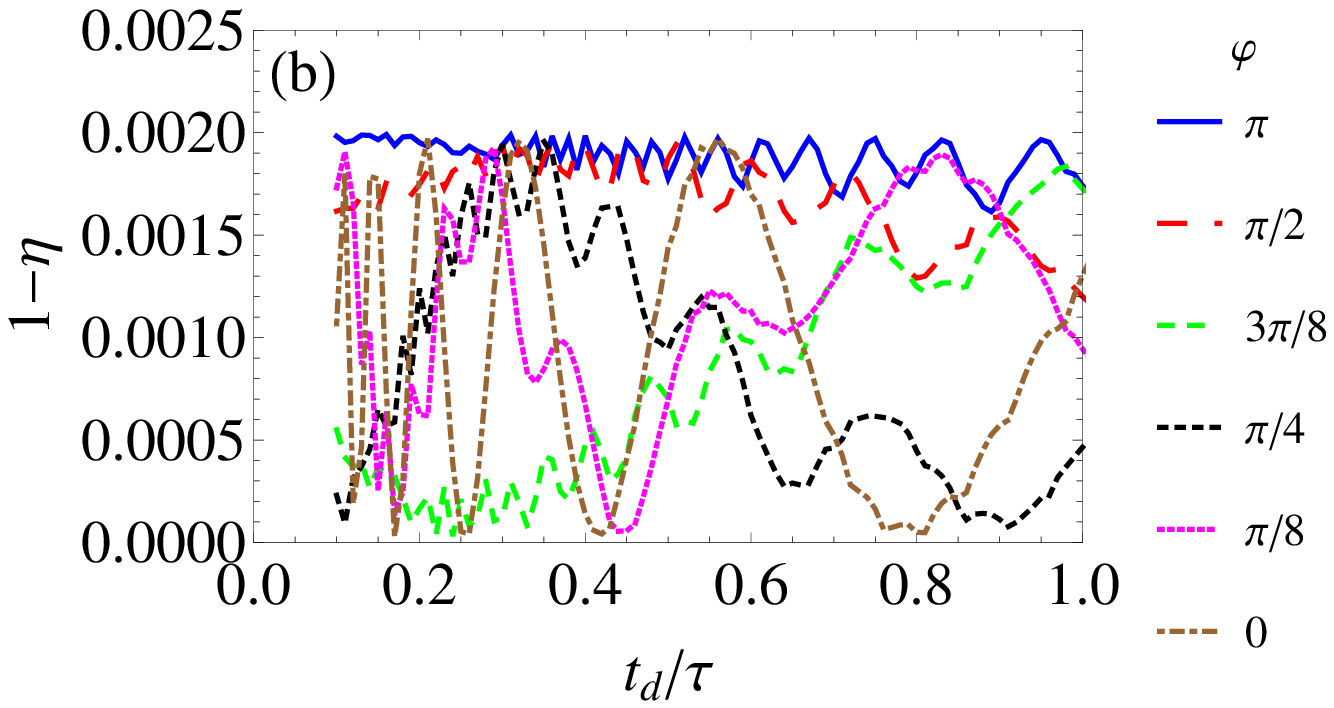}
\caption{(a) Dependence of the inefficiency $1-\eta$ on the
normalized delay $t_{\rm d}/\tau$ due to the round trip along the
transmission line, for the design efficiency $\eta_{\rm d}=0.999$
and several values of the phase shift $\varphi$ accumulated in this
round trip. The kinks at $t_{\rm d}/\tau=13.8/n$ correspond to the
integer number $n$ of the round trips within the procedure time
$t_{\rm f}$. (b) The same as in (a) for a smaller range of $t_{\rm
d}/\tau$ (the results for $t_{\rm d}/\tau <0.1$ were not
calculated). Notice that the inefficiency accounting for multiple
reflections does not exceed twice the design inefficiency, $1-\eta
\leq 2(1-\eta_{\rm d})$.
 } \label{mr1}
\end{figure}

Figure \ref{mr1} shows the numerically calculated inefficiency
$1-\eta$ of the state transfer as a function of the round-trip delay
time $t_{\rm d}$, normalized by the buildup/leakage time $\tau_{\rm
e}=\tau_{\rm r}=\tau$. Different curves represent different values
of the phase $\varphi$. The design efficiency is $\eta_{\rm
d}=0.999$. (In the simulations we also used $\omega_0/2\pi =6$ GHz,
and $\textbf{t}_{\rm e,max}=\textbf{t}_{\rm r,max}=0.05$; however,
the presented results do not depend on these parameters). We see
that the inefficiency shows an oscillatory behavior as a function of
the delay time, but it is always within the range $0\leq 1-\eta \leq
2(1-\eta_{\rm d})$. This important fact was proved in Ref.\
\cite{Kor11} in the following way. In the case with the circulator,
the losses are $1-\eta_{\rm d}=l_G^{\rm circ}+l_F^{\rm circ}$, where
$l_G^{\rm circ}=|G^{\rm circ}(t_{\rm f})|^2$ is due to the
untransmitted field [we assume here $G(0)=1$] and $l_F^{\rm circ}$
is the dimensionless energy carried away by the reflected field
$F^{\rm circ}(t)$. In the case without circulator, we can simply add
the multiple reflections of the field $F^{\rm circ}(t)$ to the
evolution with the circulator. At the final time $t_{\rm
f}$ the field $F^{\rm circ}(t)$ will linearly contribute to
$B(t_{\rm f})$, $G(t_{\rm f})$, and the field within the
transmission line [$F(t)$ for $t_{\rm f}-t_{\rm d} \leq t\leq t_{\rm
f}$]. In the worst-case scenario the whole energy $l_F^{\rm circ}$
is added in-phase to the untransmitted field $G^{\rm circ}(t_{\rm
f})$, resulting in $1-\eta = (\sqrt{l_G^{\rm circ}}+\sqrt{l_F^{\rm
circ}})^2$. Since $(\sqrt{l_G^{\rm circ}}+\sqrt{l_F^{\rm
circ}})^2\leq 2 (l_G^{\rm circ}+l_F^{\rm circ})$ always, we obtain
the upper bound for the inefficiency, $1-\eta \leq 2(1- \eta_{\rm
d})$. The lower bound $1-\eta\geq 0$ is obvious. Figure \ref{mr1}
shows that both bounds can be reached (at least approximately) with
multiple reflections at certain values of $t_{\rm d}/\tau$ and
$\varphi$ (this fact is not obvious and is even somewhat
surprising).

The dependence $\eta (t_{\rm d})$ shown in Fig.\ \ref{mr1} is quite
complicated and depends on the phase $\varphi$. We show only phases
$0\leq \varphi \leq \pi$, while for $\pi\leq \varphi \leq 2\pi$ the
results can be obtained from the symmetry $\eta (t_{\rm
d},\varphi)=\eta (t_{\rm d},2\pi-\varphi)$. As we see from Fig.\
\ref{mr1}, the oscillations of $\eta (t_{\rm d})$ generally decrease
in amplitude when $t_{\rm d}/\tau \rightarrow 0$, so that we expect
a saturation of the dependence at $t_{\rm d}/\tau \rightarrow 0$.
The exception is the case $\varphi =0$, when the oscillation
amplitude does not significantly decrease at small $t_{\rm d}/\tau$
(numerical simulations become increasingly more difficult at smaller
$t_{\rm d}/\tau$). This can be understood as due to the fact that for $\varphi
=0$ the transmission line is a resonator, which is resonant with the
frequency $\omega_{\rm e}=\omega_{\rm r}$ of the resonators.

Note that for an experiment with on-chip state transfer between
superconducting resonators, the round-trip delay time $t_{\rm d}$ is
comparable to $\omega_{\rm e(r)}^{-1}$ and therefore much smaller
than $\tau$, $t_{\rm d}/\tau \sim 10^{-2}$. This regime is outside
of the range accessible to our direct simulation method, which works
well only when $t_{\rm d}/\tau \agt 10^{-1}$. Nevertheless, we
expect that the results presented in Fig.\ \ref{mr1}(b) can be
approximately used in this case as well, because of the apparent
saturation of $\eta (t_{\rm d})$ at $t_{\rm d}\rightarrow 0$, except
when the phase $\varphi$ is close to zero.

The most important result of this section is that multiple reflections cannot increase the inefficiency $1-\eta$ by more than
twice compared with the design inefficiency $1-\eta_{\rm d}$ (as
obtained analytically and confirmed numerically).

\section{Mismatch of the resonator frequencies}\label{fre-shift}

The main idea of the state transfer protocol analyzed in this paper
is to use destructive interference to suppress the
back-reflection into the transmission line, thus providing a
high-efficiency transfer. This is why it is crucial that the
emitting and receiving resonators have almost the same frequency.
Therefore, a mismatch between the two resonator frequencies should
strongly decrease the transfer efficiency. In this section we
analyze the effect of the frequency mismatch using two models.
First, we assume a constant-in-time mismatch. Second, we consider
the time-dependent detuning of the resonator frequencies due to the
changing transmission amplitudes of the couplers, which lead to a
changing complex phase of the reflection amplitudes (see Appendix B)
and thus to the resonator frequency change.

\subsection{Constant in time frequency mismatch}

We first consider the case when the two resonator frequencies are
slightly different, $\Delta\omega \equiv \omega_{\rm e}-\omega_{\rm
r}\neq 0$, and they do not change in time. Everything else is
assumed to be ideal. It is easy to understand the effect of detuning
by using the evolution equations (\ref{lg})--(\ref{eq-A}) and
choosing $\omega_0=\omega_r$, so that $\Delta\omega_{\rm e}
=\Delta\omega$ and $\Delta\omega_{\rm r} =0$. Then, compared with
the case $\Delta\omega=0$, the emitting resonator field $G(t)$
acquires the phase factor $e^{-i\Delta\omega t}$; the same phase
factor is acquired by the transmitted field $A(t)$ in Eq.\
(\ref{le}), and this changing phase destroys the perfect phase
synchronization between $A(t)$ and $B(t)$ that is needed to cancel
the back-reflection.

\begin{figure}[t]
\includegraphics[width=8cm]{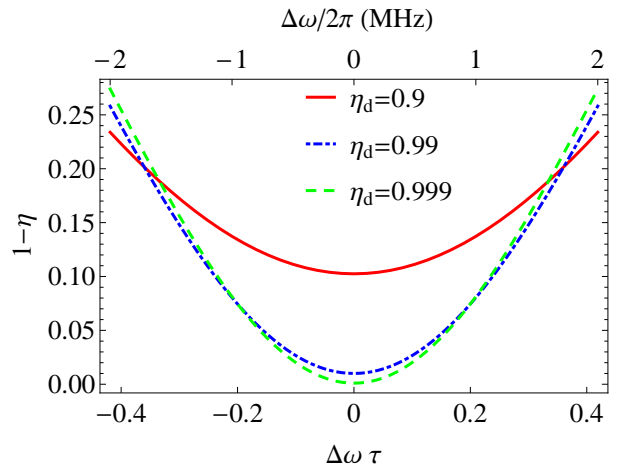}
  \caption{Inefficiency $1-\eta$ as a function of normalized detuning
$\Delta\omega\,\tau$ (lower horizontal axis) for three design
efficiencies, $\eta_{\rm d}=0.9,$ $0.99$, and $0.999$. The upper
horizontal axis shows the unnormalized detuning $\Delta\omega/2\pi$
in MHz, using the values $\omega_{\rm}/2\pi=6$ GHz and
$|\textbf{t}_{\rm e, max}|=|\textbf{t}_{\rm r, max}|=0.05$, so that
$\tau=33.3$ ns.
  } \label{cf}
\end{figure}

The numerically calculated inefficiency $1-\eta$ is shown in Fig.\
\ref{cf} as a function of the detuning $\Delta \omega$, normalized
by the inverse buildup/leakage time $\tau^{-1}$ (we assumed
$\tau_{\rm e}=\tau_{\rm r}=\tau$). We show the lines for the design
inefficiencies $\eta_{\rm d}=0.9$, 0.99, and 0.999. The results do
not depend on $\omega_{\rm r}$ and $|\textbf{t}_{\rm e(r), max}|$.
However, to express $\Delta\omega/2\pi$ in MHz on the upper
horizontal axis, we use a particular example of $\omega_{\rm
r}/2\pi=6$ GHz and $|\textbf{t}_{\rm e(r), max}|=0.05$, for which
$\tau=33.3$ ns (as in Fig.\ \ref{pulse-shapes}).

For small $|\Delta\omega \, \tau|$ and $\eta_{\rm d} \approx 1$, the
additional inefficiency due to frequency mismatch can be fitted as
\begin{equation}\label{ef}
  -\delta\eta\approx c_{\rm fm} \, (\Delta \omega\, \tau)^2, \,\,\,
  c_{\rm fm} \approx 2.
\end{equation}
For smaller $\eta_{\rm d}$ the coefficient $c_{\rm fm}$ decreases,
so that $c_{\rm fm}\approx 1.94$ for $\eta_{\rm d}=0.999$, $c_{\rm
fm}\approx 1.68$ for $\eta_{\rm d}=0.99$, and $c_{\rm fm}\approx
0.81$ for $\eta_{\rm d}=0.9$.

It is interesting that the value $c_{\rm fm}=2$ for $\eta_{\rm
d}\approx 1$ exactly coincides with the estimate derived in Ref.\
\cite{Kor11}, which we rederive here. Comparing the case
$\Delta\omega\neq 0$ with the ideal case $\Delta\omega = 0$, we can
think that $A(t)$ acquires the extra phase factor $e^{-i\Delta\omega
(t-t_{\rm m})}$, where $t_{\rm m}$ is the mid-time of the procedure
(see Fig.\ \ref{pulse-shapes}); the overall factor $e^{i\Delta\omega
t_{\rm m}}$ is not important, affecting only the final phase
$\varphi_{\rm f}$. Then we can think that at $t=t_{\rm m}$ we still
have an almost perfect cancellation of the back-reflection,
$F(t_{\rm m})\approx 0$; however, at $t\neq t_{\rm m}$ the extra
phase causes the back-reflected wave $|F(t)|\approx
|A(t)(e^{-i\Delta\omega (t-t_{\rm m})}-1)|$. Now using
$|A(t)|=|A(t_{\rm m})| e^{-|t-t_{\rm m}|/2\tau}$ and assuming
$|\Delta \omega|\tau \ll 1$ (so that we can expand the exponent in
the relevant time range), we find $|F(t)|\approx |A(t_{\rm m})| \,
e^{-|t-t_{\rm m}|/2\tau} |\Delta\omega (t-t_{\rm m})|$. Finally
integrating the loss, $\int |F(t)|^2 dt$, and normalizing it by the
transferred ``energy'' $\int |A(t_{\rm m})|^2 e^{-|t-t_{\rm
m}|/\tau} dt$, we obtain the added inefficiency $-\delta \eta
\approx 2\, (\Delta\omega\, \tau)^2$.

Using this derivation, it is easy to understand why the coefficient
$c_{\rm fm}$ in Eq.\ (\ref{ef}) decreases with decreasing $\eta_{\rm
d}$. This occurs because the integration of $|F(t)|^2$ is limited by
the range $0<t<t_{\rm f}=-2\tau\ln (1-\eta_{\rm d})$, which becomes
shorter for smaller $\eta_{\rm d}$. Thus we can estimate $c_{\rm
fm}$ as $c_{\rm fm}\approx \int_0^{-\ln (1-\eta_{\rm d})}x^2 e^{-x}
dx = 2 - (1-\eta_{\rm d})[2-2\ln (1-\eta_{\rm d})+\ln^2 (1-\eta_{\rm
d})]$, which fits the numerical results very well.

As expected, even small detuning significantly decreases the
transfer efficiency. For example, to keep the added inefficiency
under 1\%, $-\delta\eta <0.01$, we need the detuning to be less than
0.4 MHz in the above example ($\tau =33.3$ ns), which is not very easy to
achieve in an experiment.

\subsection{Time-dependent detuning due to changing coupling}
\label{sec-variable-detuning}

In an actual experimental coupler, the parameters are interrelated,
and a change of the coupling strength by varying $|\textbf t|$ may
lead to a change of other parameters. In particular, for the coupler
realized experimentally in Refs.\ \cite{Yin13,Wen14}, the change of
$|\textbf t|$ causes a small change of the complex phases of the
transmission and reflection amplitudes $\textbf t$ and $\textbf
r^{\rm in (out)}$. The phase change of $\textbf r^{\rm in}$ (from
the resonator side) causes a change of the resonator frequency.
Thus, changing the coupling causes the frequency detuning, as was
observed experimentally \cite{Yin13}. Since the frequency mismatch
between the two resonators strongly decreases the efficiency of the
state transfer, this is a serious problem for the protocol discussed
in our paper. Here we analyze this effect quantitatively and discuss
with which accuracy the detuning should be compensated (e.g.\ by
another tunable element) to preserve the high-efficiency transfer.

Physically, the resonator frequency changes because the varying coupling changes the boundary condition at the end of the coplanar waveguide resonator (see Fig.\ \ref{tuna} in Appendix B). Note that a somewhat similar frequency change due to changing coupling with a transmission line was studied in Ref.\ \cite{Goppl-08}.

As discussed in Appendix B, if we use the tunable couplers of Ref.\
\cite{Yin13,Wen14}, then the transmission and refection amplitudes
$\textbf t_j$ and $\textbf r_j^{\rm in (out)}$ for the two
resonators ($j=\rm e,\, r$) are given by the formulas
    \begin{align}
& \textbf{t}_{j}=-i\frac{2\omega_{j} M_{j}}{1+b_{j}}
\sqrt{\frac{R_j}{R_{\rm
tl}}}\left(\frac{1}{R_{j}}+\frac{-ib_{j}}{\omega _{j}
L_{e,j}}\right)\frac{1}{1-i\omega_{j} L_{2,j}/R_{\rm tl}},
  \label{ttn} \\
  & \textbf{r}_{j}^{\rm in}=-\frac{1-b_{j}}{1+b_{j}}, \,\,\,
  \textbf{r}_{j}^{\rm out}= -(\textbf{r}_{j}^{\rm in})^*
  \frac{\textbf{t}_{j}}{\textbf{t}_{j}^*} ,
  \label{rrn}
 \end{align}
where
\begin{equation}\label{b-def}
  b_j = \frac{i\omega L_{1,j}/R_{j}}{\displaystyle
\frac{L_{1,j}}{L_{e,j}}+\left[ 1-\frac{i\omega_j M_j^2}{R_{\rm
tl}L_{1,j}(1+i\omega_j L_{2,j}/R_{\rm tl})}\right]^{-1}} ,
\end{equation}
$M_j$ is the effective mutual inductance in $j$th coupler (the main
tunable parameter controlled by magnetic flux in the SQUID loop),
$R_{j}$ and $R_{\rm tl}$ are the wave impedances of the resonators
and the transmission line, $\omega_j$ are the resonator frequencies,
and $L_{1,j}$, $L_{2,j}$, and $L_{e,j}$ are the effective
inductances used to describe the coupler (see details in Appendix
B). Note that Eqs.\ (\ref{ttn}) and (\ref{b-def}) are slightly
different from the equations in the Supplementary Information of
Ref.\ \cite{Yin13} and the derivation in Appendix B: the difference
is that the imaginary unit $i$ is replaced with $-i$ to conform with
the chosen rotating frame definition $e^{-i\omega t}$ in Eqs.\
(\ref{lg}) and (\ref{le}).

For the typical experimental parameters, $|b_j|\ll 1$, so that
$\textbf{r}_{j}^{\rm in}\approx -1$, while $\textbf{t}_{j}$ is mostly
imaginary. Note that $\omega_{\rm e}\approx \omega_{\rm r}\approx
\omega_{0}$, so in Eqs.\ (\ref{ttn}) and (\ref{b-def}) we can
replace $\omega_j$ with $\omega_0$. Also note that there is no
coupling, $\textbf{t}_{j}=0$, when $M_j=0$, and the coupling changes
sign when $M_j$ crosses zero.

Tuning $M_j$, we control $|\textbf{t}_{j}|$. However, the complex
phase of $\textbf{t}_{j}$ slightly changes with changing $M_j$
because $b_j$ in Eq.\ (\ref{ttn}) depends on $M_j$ and also
$L_{1,j}$ and $L_{2,j}$ depend on $M_j$ -- see Appendix B. Changing
the phase of $\textbf{t}_{j}$ leads to the phase mismatch in the
state transfer protocol, degrading its efficiency. However, this is
a relatively minor effect, while a much more serious effect is the
dependence of the complex phase of $\textbf{r}_{j}^{\rm in}$ on
$M_j$ via its dependence on $b_j$ in Eq.\ (\ref{rrn}), leading to
the resonator frequency change.

For the rotating frame $e^{-i\omega t}$ and quarter-wavelength
resonator (which we assume here) the change $\delta (\arg
\textbf{r}_{j}^{\rm in})$ of the phase of $\textbf{r}_{j}^{\rm in}$
changes the resonator frequency by
    \be
\delta \omega_j \approx -(\omega_0/\pi) \, \delta (\arg
\textbf{r}_{j}^{\rm in}),
    \label{delta-omega}\ee
where we used $\omega_j\approx \omega_0$. Assuming for simplicity
that the resonators are exactly on resonance ($\omega_{\rm
e}=\omega_{\rm r}=\omega_0$) when there is no coupling ($M_{\rm
e}=M_{\rm r}=0$), we can write the variable detunings to be used in
the evolution equations (\ref{lg}) and (\ref{le}) as
    \be
    \Delta \omega_j = \omega_j -\omega_0= -\frac{\omega_0}{\pi}
    \left[ \arg \textbf{r}_{j}^{\rm in}(M_j)-
    \arg \textbf{r}_{j}^{\rm in}(0)  \right],
    \label{Delta-omega-j}\ee
where $\textbf{r}_{j}^{\rm in}(M_j)$ describes dependence on $M_j$.
Since $|\textbf{t}_{j}|$ also depends on $M_j$ (linearly to first
approximation), we have an implicit dependence $\Delta
\omega_j (|\textbf{t}_{j}|)$, which is linear for small
$|\textbf{t}_{j}|$ [see Eq.\ (\ref{Delta-omega-ratio}) in Appendix
B] and becomes nonlinear for larger $|\textbf{t}_{j}|$ .

\begin{figure}[t]
\includegraphics[width=8cm]{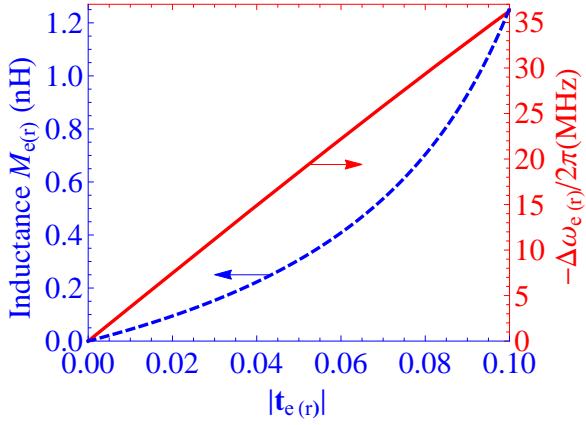}
 \caption{Red solid line: the resonator frequency detuning
$-\Delta\omega_{\rm e(r)}/2\pi$ caused by changing $|\textbf{t}_{\rm
e(r)}|$ for a particular set of parameters of the coupler (see
text). Blue dashed line: the corresponding value of the coupler
mutual inductance $M_{\rm e(r)}$. The arrows indicate the corresponding vertical axes.
    } \label{Mvst}
\end{figure}

\begin{figure}[t]
  \includegraphics[width=8cm]{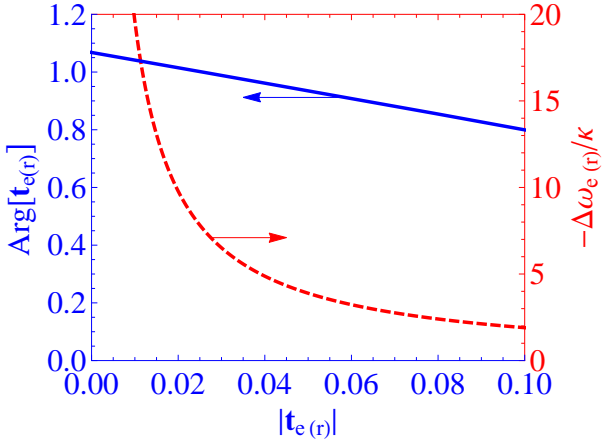}
\caption{Solid line: the phase of the transmission amplitude, $\arg
(\textbf{t}_{\rm e(r)})$, as a function of its absolute value
$|\textbf{t}_{\rm e(r)}|$ for a particular set of coupler parameters
(see text). Dashed line: the normalized detuning $-\Delta
\omega_{\rm e(r)}/\kappa_{\rm e(r)}=-\pi \Delta \omega_{\rm
e(r)}/\omega_{\rm e(r)}|\textbf{t}_{\rm e(r)}|^2$.
   }
   \label{argt}   \end{figure}

This dependence $\Delta \omega_{\rm e(r)} (|\textbf{t}_{\rm e(r)}|)$
is shown in Fig.\ \ref{Mvst} by the solid line  for the parameters
of the coupler similar (though not equal) to the parameters of the
experimental coupler \cite{Yin13}: $R_{\rm e(r)}=80 \, \Omega$,
$R_{\rm tl}=50\, \Omega$, $\omega_0/2\pi= 6\, \rm GHz$,
$L_{1,j}-M_j=L_{2,j}-M_j=620\,\rm pH$, and $L_{e,j}=180 \, \rm pH$
(see Appendix B). In particular, Fig.\ \ref{Mvst} shows that
$|\textbf{t}_{\rm e(r)}|=0.05$ corresponds to the frequency change
by $-18.6$ MHz, which is a very big change compared to what is
tolerable for a high-efficiency state transfer (see Fig.\ \ref{cf}).
The same detuning normalized by $\kappa_{\rm e(r)}=|\textbf{t}_{\rm
e(r)}|^2\omega_{\rm e(r)}/\pi$ is shown in Fig.\  \ref{argt} by the
dashed line.

The value of $M_{\rm e(r)}$ needed to produce a given
$|\textbf{t}_{\rm e(r)}|$ is shown in Fig.\ \ref{Mvst} by the dashed
line. It is interesting that the dependence $M(|\textbf{t}|)$ is
significantly more nonlinear than the dependence  $\Delta \omega
(|\textbf{t}|)$, indicating that the nonlinearities of $|\textbf{t}
(M)|$ and $\Delta\omega (M)$ in Eqs.\ (\ref{ttn}) and
(\ref{Delta-omega-j}) partially cancel each other (see Appendix B).

The solid line in Fig.\ \ref{argt} shows dependence of the phase
$\arg [\textbf{t}_{\rm e(r)}]$ on the absolute value $|
\textbf{t}_{\rm e(r)}|$. Even though the phase change looks significant, it produces a relatively minor decrease in the protocol
inefficiency (as we will see later) because the loss is quadratic in
the phase mismatch.

We numerically simulate the state transfer protocol, accounting for
the frequency change of the resonators and phase change of
$\textbf{t}_{\rm e (r)}$ in the following way. First, we use the
ideal pulse shapes $|\textbf{t}_{\rm e}(t)|$ and $|\textbf{t}_{\rm
r}(t)|$ from Eqs.\ (\ref{tem})--(\ref{trec-2}), assuming a symmetric
setup ($\tau_{\rm e}=\tau_{\rm r}$). Then we calculate the
corresponding dependences $M_{\rm e}(t)$ and $M_{\rm r}(t)$ using
Eq.\ (\ref{ttn}) and find $\textbf{t}_{\rm e}(t)$ and
$\textbf{t}_{\rm r}(t)$ (now with time-dependent phases) using the
same Eq.\ (\ref{ttn}), and also find the detunings $\Delta
\omega_{\rm e}(t)$ and $\Delta \omega_{\rm r}(t)$ using Eq.\
(\ref{Delta-omega-j}). After that we solve the evolution equations
(\ref{lg})--(\ref{eq-A}), neglecting multiple reflections. Note that
we convert $|\textbf{t}_{j}(t)|$ into $M_j(t)$ by first numerically
calculating $|\textbf{t}_{j}(M_{j})|$ from Eq.\ (\ref{ttn}), then
fitting the inverse dependence $M_j(|\textbf{t}_{j}|)$ with a
polynomial of 40th order, and then using this polynomial for the
conversion.

\begin{figure}[t]
\includegraphics[width=8cm]{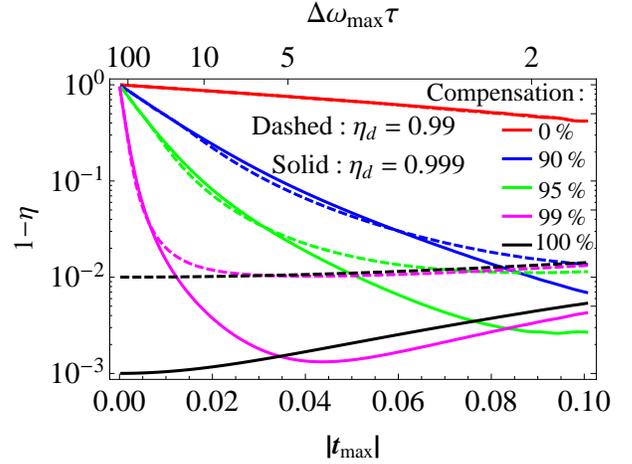}
\caption{ Inefficiency $1-\eta$ as a function of $|\textbf{t}_{\rm
e, max}|=|\textbf{t}_{\rm r, max}|$ for the couplers with parameters
described in the text. The solid lines are for the design efficiency
$\eta_{\rm d}=0.999$, while the dashed lines are for $\eta_{\rm
d}=0.99$. The red lines show the results without compensation of the
frequency detuning $\Delta \omega_{\rm e(r)}(t)$ caused by changing
$|\textbf{t}_{\rm e(r)}(t)|$ and correspondingly changing $\arg
(\textbf{r}_{\rm e(t)}^{\rm in})$. The blue lines assume 90\%
compensation of this detuning, 95\% compensation for the green
lines, 99\% compensation for the magenta lines, and full 100\%
compensation for the black lines. For the black lines the extra
inefficiency is caused only by changing phases of $\textbf{t}_{\rm
e(r)}$. The upper horizontal axis shows the product $\Delta
\omega_{\rm max} \, \tau$, corresponding to $|\textbf{t}_{\rm
max}|$.
    } \label{comp}
\end{figure}

Figure \ref{comp} shows the numerically calculated inefficiency
$1-\eta$ of the transfer protocol as a function of the maximum
transmission amplitude $|\textbf{t}_{\rm e, max}|=|\textbf{t}_{\rm
r, max}|$ for the above example of the coupler parameters and design
efficiencies $\eta_{\rm d}=0.99$ and 0.999. Besides showing the
results for the usual protocol (red lines), we also show the results
for the cases when the frequency detuning [Eq.\
(\ref{Delta-omega-j})] is reduced by a factor of 10 (90\%
compensation, blue lines), 20 (95\% compensation, green lines), 100
(99\% compensation, magenta lines) and fully eliminated (100\%
compensation, black lines). Such compensation can be done
experimentally by using another circuit element, affecting the
resonator frequency, e.g., tuning the phase of the reflection
amplitude at the other end of the resonator by a SQUID-controlled
inductance.

We see that without compensation of the frequency detuning the state
transfer protocol cannot provide a high efficiency: $\eta = 0.33$
for $|\textbf{t}_{\rm max}|=0.05$ and $\eta =0.58$ for
$|\textbf{t}_{\rm max}|=0.1$. However, with the detuning
compensation the high efficiency may be restored. As we see from
Fig.\ \ref{comp}, the state transfer efficiency above 99\% requires
the detuning compensation at least within 90\%--95\% range
(depending on $|\textbf{t}_{\rm max}|$). Note that even with 100\%
compensation, the efficiency is less than in the ideal case. This is
because of the changing phases of $\textbf{t}_{\rm e}(t)$ and
$\textbf{t}_{\rm r}(t)$. However, this effect is minor in comparison
with the effect of detuning.

It is interesting that the curves in Fig.\ \ref{comp} decrease with
increasing $|\textbf{t}_{\rm max}|$ when $|\textbf{t}_{\rm max}|$ is
not too large. This may seem counterintuitive, since larger
$|\textbf{t}|$ leads to larger detuning, and so we would naively expect
larger inefficiency at larger $|\textbf{t}_{\rm max}|$. The
numerical result is opposite because the duration of the procedure
decreases, scaling as $\tau \propto |\textbf{t}_{\rm max}|^{-2}$.
Therefore if the largest detuning scales linearly,
$|\Delta\omega_{\rm max}|\propto |\textbf{t}_{\rm max}|$, then the
figure of merit $|\Delta\omega_{\rm max}\tau |$ scales as
$|\textbf{t}_{\rm max}|^{-1}$, thus explaining the decreasing part
of the curves in Fig.\ \ref{comp}. The upper horizontal axis in
Fig.\ \ref{comp} shows $|\Delta\omega_{\rm max}\tau |$, which indeed
decreases with increasing $|\textbf{t}_{\rm max}|$ (see also the
dashed line in Fig.\ \ref{argt}).

More quantitatively, let us assume a linear detuning, $\Delta
\omega_{\rm e(r)}=k \, |\textbf{t}_{\rm e(r)}|$, where the
coefficient $k$ is given by Eq.\ (\ref{Delta-omega-ratio})
multiplied by the uncompensated fraction of the detuning. Assuming a
small deviation from the ideal protocol, the transmitted wave is
$|A(t)|=|A(t_m)|\,  e^{-|\Delta t|/2\tau}$, where $\Delta t=t-t_{\rm
m}$. At the mid-time $t_{\rm m}$ the resonator frequencies coincide,
but at $t>t_{\rm m}$ the receiving resonator frequency changes so
that $\Delta\omega = \omega_{\rm e}-\omega_{\rm r} = k(
|\textbf{t}_{\rm r}(t_{\rm m})|- |\textbf{t}_{\rm r}(t)|)$. Using
Eq.\ (\ref{trec}) we find $\Delta\omega=k \, |\textbf{t}_{\rm
max}|\, [1- (2e^{\Delta t/\tau} -1)^{-1/2}]$. The accumulated phase
mismatch is then $\phi (t)=\int_{t_m}^t \Delta \omega (t') \, dt' $,
which produces the reflected wave $|F|\approx |A \phi|$, assuming
small $\phi$. The inefficiency due to the reflected wave loss is
then $1-\eta \approx \int_{t_m}^\infty |F(t)|^2\, dt /
\int_{t_m}^\infty |A(t)|^2\, dt$ (note that due to symmetry the same
relative loss is before and after $t_{\rm m}$). Therefore $1-\eta
\approx \tau^{-1} k^2 |\textbf{t}_{\rm max}|^2 \int_0^\infty \{
\int_0^x [1- (2e^{\Delta t/\tau} -1)^{-1/2}] \, d\Delta t \}^2
e^{-x/\tau} dx$, and calculating the integral numerically we obtain
$1-\eta = 0.63 \, k^2 \tau^2 |\textbf{t}_{\rm max}|^2$ [the
numerical value of the integral is somewhat smaller than 0.63 if we
limit the outer integration by $-\tau\ln (1-\eta_{\rm d})$]. Finally
using $\tau= \pi/\omega_0 |\textbf{t}_{\rm max}|^2$, we obtain
$1-\eta \approx 0.6 \, (k\pi/\omega_0 |\textbf{t}_{\rm max}|)^2$.

Numerical results in Fig.\ \ref{comp} reproduce the scaling
$1-\nobreak \eta \propto (k/|\textbf{t}_{\rm max}|)^2$ for the
significant part of the curves for $\eta_{\rm d}=0.999$ (when
plotted in log-log scale); however, the prefactor in the numerical
fitting is somewhat different from what we obtained above: $1-\eta
\approx 0.4\, (k\pi/\omega_0 |\textbf{t}_{\rm max}|)^2$.
Note that at sufficiently large $|\textbf{t}_{\rm max}|$  the green
and red curves in Fig.\ \ref{comp} reach a minimum and then start to
increase. This occurs because the inefficiency due to changing phase
of $\textbf{t}_{\rm e(r)}$ increases with increasing
$|\textbf{t}_{\rm max}|$, in contrast to the effect of frequency
detuning.

Actually, our analysis of the transfer process in the case of
complete compensation of detuning is not fully accurate. The reason
is that in the evolution equations (\ref{lg})--(\ref{eq-A}) we took
into account the frequency change due to changing $\textbf{r}^{\rm
in}_{\rm e(r)}$, but we did not take into account another (very
small) effect due to changing $\textbf{r}^{\rm in}_{\rm e(r)}$. It
is easy to understand the origin of this effect in the following
way. There is a phase difference $\arg(\textbf{r}^{\rm in}_{\rm r})$
between the field $B$ propagating away from the transmission line
and the similar field propagating towards the transmission line [see
Eq.\  (\ref{F-form1}) and discussion below it]. Changing
$\arg(\textbf{r}^{\rm in}_{\rm r})$ alters this phase difference,
thus affecting both fields and correspondingly leading to an
extra term, neglected in Eq.\ (\ref{le}). Similarly, changing
$\arg(\textbf{r}^{\rm in}_{\rm e})$ leads to an extra term in Eq.\
(\ref{lg}) for $G$. However, as can be seen from Fig.\ \ref{Mvst}
and Eq.\ (\ref{delta-omega}), the change of $\arg(\textbf{r}^{\rm
in}_{\rm e(r)})$ is less than 0.02 for $|\textbf{t}_{\rm e(r)}|$
varying between 0 and 0.1, which is much less than the change of
$\arg(\textbf{t}_{\rm e(r)})$ in Fig.\ \ref{argt}. Therefore, the
neglected effect is much less than the effect due to changing
$\arg(\textbf{t}_{\rm e(r)})$, which by itself is almost negligible,
as seen in Fig.\ \ref{comp}. Note that the compensation for changing
phases can be done experimentally in the same way as the
compensation for the detuning, so that in principle the efficiency
decrease analyzed in this section can be fully avoided.

Overall, we see that the detuning of the resonator frequencies due
to a changing coupling is a serious problem for the state transfer
protocol. A high-efficiency state transfer is possible only with
additional experimental effort to compensate for this detuning. The
required compensation accuracy is crudely within 90\%--95\% range.
The use of a shorter protocol (by using a stronger coupling) helps
to increase the efficiency. Note that the frequency compensation is done routinely for the tunable coupler of Refs.\ \cite{Hoffman-11,Sri14}; similarly, the phase compensation can be naturally realized in the tunable coupler of Refs.\ \cite{Pechal2014,Zeytinoglu2015}.

\section{Conclusion}\label{con}

In this paper we have analyzed the robustness of the quantum state
transfer protocol of Ref.\ \cite{Kor11} for the transfer between two
superconducting resonators via a transmission line. The protocol is
based on destructive interference, which cancels
the back-reflection of the field into the transmission line at the
receiving end (we believe this explanation is more natural than the terminology of time reversal, introduced in Ref.\ \cite{Cir97}). This is achieved by using tunable couplers for both
resonators and properly designed time-dependences (pulse
shapes) of the transmission amplitudes $\textbf{t}_{\rm e}(t)$ and
$\textbf{t}_{\rm r}(t)$ for these couplers. Nearly-perfect transfer
efficiency $\eta$ can be achieved in the ideal case. We have focused
on analyzing additional inefficiency due to deviations from the
ideal case.

The ideal pulse shapes of the transmission amplitudes [Eqs.\
(\ref{tem})--(\ref{trec-2})] depend on several parameters; we have
studied additional inefficiency due to deviations of these
parameters from their design values. Below, we summarize our results
by presenting the tolerable deviations for a fixed additional
inefficiency of $-\delta\eta=0.01$ (because of quadratic scaling,
the tolerable inaccuracies for $-\delta\eta=0.001$ are about 3.2
times smaller). For the relative deviations of the maximum
transmission amplitudes $|\textbf{t}_{\rm e, max}|$ and
$|\textbf{t}_{\rm r, max}|$, the tolerable ranges are $\pm 10\%$ if
only one of them is changing and $\pm 5\%$ if both of them are
changing simultaneously [see Fig.\ \ref{tmax} and  Eq.\
(\ref{bothtm})]. For the relative deviations of the time scale
parameters $\tau_{\rm e}$ and $\tau_{\rm r}$ describing the
exponential increase/decrease of the transmitted field, the
tolerable ranges are $\pm 17\%$ if only one of them is changing and
$\pm 11\%$ if both of them are changing simultaneously [see Fig.\
\ref{fig-tau} and  Eq.\ (\ref{tauboth})]. For the mismatch between
the mid-times $t_{\rm m}$ of the procedure in the two couplers, the
tolerable range is $\pm 0.2\tau\simeq \pm 6$ ns [see Fig.\
\ref{figtm} and  Eq.\ (\ref{tm})]. For a nonlinear distortion
described by warping parameters $\alpha_{\rm e}$ and $\alpha_{\rm
r}$ [see Eq.\ (\ref{tnw})], the tolerable parameter range is $\pm
0.2$ if the distortion affects only one coupler and $\pm 0.13$ if
the distortion affects both couplers. Our results show that
smoothing of the pulse shapes by a Gaussian filter practically does
not affect the inefficiency; even filtering with the width
$\sigma\simeq \tau\simeq 30$ ns is still tolerable. When the pulse
shapes are distorted by an additional (relatively high-frequency)
noise, the tolerable range for the standard deviation of
$|\textbf{t}_{\rm e(r)}|$ is 7\% of the instantaneous value and 3\%
of the maximum value [see Fig.\ \ref{noise} and  Eq.\
(\ref{noise-fit})]. Overall, we see that the state transfer
procedure is surprisingly robust to various distortions of the pulse
shapes.

We have also analyzed the effect of multiple reflections and found
that it can both increase or decrease the transfer efficiency.
However, even in the worst case, this effect cannot increase the
inefficiency $1-\eta$ by more than a factor of 2 (see Fig.\
\ref{mr1}). The energy dissipation in the transmission line or in
the resonators can be a serious problem for the state transfer
protocol. The description of the effect is simple [see Eq.\
(\ref{dissip})]; for a high-efficiency transfer we can tolerate only
a weak dissipation  $1-\eta_{\rm tl}$ in the transmission line, and
we also need the procedure duration $t_{\rm f}$ to be much shorter
than the energy relaxation time $T_1$. In particular, for
$-\delta\eta =0.01$ we need $\eta_{\rm tl}>0.99$ and $T_1>100 \,
t_{\rm f}$.

The major problem in realizing the state transfer protocol
is the frequency mismatch between the two resonators, since the
destructive interference is very sensitive to the frequency
mismatch. For a fixed detuning, the tolerable frequency mismatch
$(\omega_{\rm e}-\omega_{\rm r})/2\pi$ for $-\delta\eta =0.01$ is
only $\pm 0.01/\tau \simeq \pm 0.4$ MHz [see Fig.\ \ref{cf} and Eq.\
(\ref{ef})]; the tolerable range is a factor of $\sqrt{10}$ smaller
for $-\delta\eta =0.001$.  An even more serious problem is the
change of the resonator frequencies caused by changing couplings,
which for the coupler of Ref.\ \cite{Yin13} is on the order of 20
MHz [see Fig.\ \ref{Mvst} and Eq.\ (\ref{Delta-omega-ratio}) in
Appendix B]. Without active compensation for this frequency change,
a high-efficiency state transfer is impossible. Our numerical
results show (see Fig.\ \ref{comp}) that to realize efficiency $\eta
=0.99$, the accuracy of the compensation should be at least 90\%
(i.e., the frequency change should be decreased by an order of
magnitude). It is somewhat counterintuitive that a better efficiency
can be obtained by using a higher maximum coupling, which increases
the frequency mismatch but decreases duration of the procedure (see
Fig.\ \ref{comp}). Another effect that decreases the efficiency is
the change of the phase of the transmission amplitude with changing
coupling. However, this effect produces a relatively minor decrease
of the efficiency (see Fig.\ \ref{comp}).

In most of the paper we have considered a classical state transfer,
characterized by the (energy) efficiency $\eta$. However, all the
results have direct relation to the transfer of a quantum state (see
Appendix A). In particular, for a qubit state transfer, the quantum process fidelity
$F_\chi$ is $F_\chi \approx 1-(1-\eta)/2$ for $\eta \approx 1$ [see
Eq.\ (\ref{pfn})].

The quantum state transfer protocol analyzed in this paper has
already been partially realized experimentally. In particular, the
realization of the proper (exponentially increasing) waveform for
the quantum signal emitted from a qubit has been demonstrated in
Ref.\ \cite{Sri14} (a reliable frequency compensation has also been demonstrated in that paper). The capture of such a waveform with 99.4\%
efficiency has been demonstrated in Ref.\ \cite{Wen14}. We hope that
the full protocol that combines these two parts will be realized in
the near future.

\begin{acknowledgments}
  The authors thank Kyle Keane who was significantly
involved in this work at the initial stage. We also thank Justin
Dressel, Mostafa Khezri, James Wenner, Andrew Cleland, and John
Martinis for useful discussions, and thank Justin Dressel for proofreading the manuscript.
  The research was funded by the
Office of the Director of National Intelligence (ODNI), Intelligence
Advanced Research Projects Activity (IARPA), through the Army
Research Office Grant No. W911NF-10-1-0334. All statements of fact,
opinion or conclusions contained herein are those of the authors and
should not be construed as representing the official views or
policies of IARPA, the ODNI, or the U.S. Government. We also
acknowledge support from the ARO MURI Grant No. W911NF-11-1-0268.
\end{acknowledgments}

\appendix

\section{Quantum state transfer using the beam splitter theory}

In this Appendix we discuss the {\it quantum} theory of state
transfer using the optical language of beam splitters. The starting
point is Eq.\ (\ref{Bf-general}), in which the resulting classical
field $B(t_{\rm f})$ has the contribution $\sqrt{\eta} \,
e^{i\varphi_{\rm f}} G(0)$ from the transferred field $G(0)$ and
also contributions from other fields. This equation describes a
unitary transformation, which can be modeled as a result of adding
the (infinite number of) fields [$B(0)$ and time-binned $V(t)$] by using a system of (infinite
number of) beam splitters. Then using linearity of the evolution, we
can simply replace the classical fields with the corresponding
annihilation operators for quantum fields, thus developing the
quantum theory of the state transfer.

In the case when all other fields in Eq.\ (\ref{Bf-general}) except
$G(0)$ correspond to vacuum, it is sufficient to consider one beam
splitter because a linear combination of vacua is still vacuum.
This is why in this Appendix we mainly discuss one beam splitter
(characterized by the amplitude $\sqrt{\eta}$ and phase
$\varphi_{\rm f}$ in the main path), with
the initial state to be transferred at one arm and vacuum state at the
other arm. Note that notations in this Appendix are different from
the notations in the main text.

  \begin{figure}[t]
\includegraphics[width=5cm]{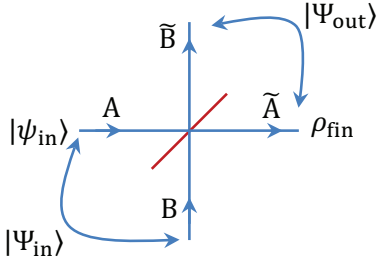}
  \caption{
A beam splitter with input classical fields $A$ and $B$ transformed
into the output fields $\tilde A$ and $\tilde B$, with the main
transformation $A\rightarrow \tilde A$ characterized by the
amplitude $\sqrt{\eta}$ and phase shift $\varphi_{\rm f}$. In the quantum
formulation the input state $|\Psi_{\rm in}\rangle$ is transformed
into the output state $|\Psi_{\rm out}\rangle$. In particular, in
Sec.\ \ref{sec-ap-arb} we consider the input state $|\Psi_{\rm
in}\rangle=|\psi_{\rm in}\rangle |0\rangle$, calculate $|\Psi_{\rm
out}\rangle$, and then reduce it to the density matrix $\rho_{\rm
fin}$ of the main output arm, by tracing over the other output arm.
   }\label{bsp}
\end{figure}

Let us start with revisiting the quantum theory of a beam splitter
\cite{Ger06}  (Fig.\ \ref{bsp}). The quantum theory follows the classical description of the beam splitter, which is characterized
by the following relations between the input classical fields $A$
and $B$, and the output classical fields $\tilde A$ and $\tilde B$:
    \begin{eqnarray}
&& \tilde A = \sqrt{\eta} \, e^{i\varphi_1} A + \sqrt{1-\eta} \,
e^{i\varphi_2}B,
    \label{tilde-A}
    \\
&& \tilde B =\sqrt{\eta} \, e^{i\varphi_3} B -\sqrt{1-\eta} \,
e^{i(\varphi_1-\varphi_2+\varphi_3)} A,
    \label{tilde-B}\end{eqnarray}
where $\varphi_1=\varphi_{\rm f}$ and other phases are introduced to
describe a general unitary transformation (these phases can include
phase shifts in all four arms). Exactly the same relations also apply in the quantum case for the annihilation operators $\tilde a$ and $\tilde b$ of
the fields at the output arms and the annihilation operators $a$ and $b$ of the fields at the input arms.

In general, we want to find an output quantum state $|\Psi_{\rm
out}\rangle$ for a given input state $|\Psi_{\rm in}\rangle$, which in principle can be an entangled state of the two input modes. This can
be done \cite{Ger06} by applying the following steps:
\begin{enumerate}
\item Express the input state $|\Psi_{\rm in}\rangle$ in terms of
the input creation operators $a^\dagger$ and $b^\dagger$, and
vacuum.
\item Using Eqs.\ (\ref{tilde-A}) and (\ref{tilde-B}), express $A$
and $B$ via $\tilde A$ and $\tilde B$. These are the equations
expressing $a$ and $b$ in terms of $\tilde a$ and $\tilde b$.
Conjugate these equations to express $a^\dagger$ and $b^\dagger$ in
terms of $\tilde a^\dagger$ and $\tilde b^\dagger$.
\item Substitute the operators $a^\dagger$ and $b^\dagger$ used
in the step 1 by their expressions in terms of $\tilde a^\dagger$
and $\tilde b^\dagger$ obtained in step 2. This substitution
gives $|\Psi_{\rm out}\rangle$.
\end{enumerate}

Now let us apply this substitution method to find the resulting
state in the receiving resonator when an arbitrary quantum state is
transferred from the emitting resonator.

\subsection{Transfer of an arbitrary quantum state}
\label{sec-ap-arb}

Let us assume that the initial state $|\psi_{\rm in}\rangle$ in the
emitting resonator is
    \be
|\psi_{\rm in}\rangle = \sum_{n}\alpha_{n}|n\rangle =
\sum_{n}\frac{\alpha_n (a^{\dagger})^{ n}}{\sqrt{n!}}|0\rangle ,
\,\,\, \sum_n |\alpha_n|^2 =1,
   \label{psi-in}\ee
while all other fields involved in the transfer procedure are
vacua (in particular, this assumes zero temperature). Then the
two-arm input state $|\Psi_{\rm in}\rangle$  for the beam splitter
is the same, except the vacuum $|0\rangle$ in Eq.\ (\ref{psi-in}) is
now understood as the vacuum $|\bf{0}\rangle$ for all possible modes.

The transfer procedure is characterized only by the efficiency
$\eta$ and the phase $\varphi_{\rm f}=\varphi_1$, while other phases
$\varphi_{2}$ and $\varphi_3$ in Eqs.\ (\ref{tilde-A}) and
(\ref{tilde-B}) are undefined. However, even though the resulting
state $|\Psi_{\rm out}\rangle$ will depend on $\varphi_{2}$ and
$\varphi_{3}$, the resulting density matrix $\rho_{\rm fin}$,
obtained from $|\Psi_{\rm out}\rangle$  by tracing over the other
output arm, will not depend on $\varphi_{2}$ and $\varphi_{3}$. This
is because arbitrary $\varphi_2$ and $\varphi_3$ can be produced by
placing phase shifters in the ancillary input and output arms
($B$-arm and $\tilde B$-arm in Fig.\ \ref{bsp}); shifting the phase
of vacuum in the $B$-arm does not produce any effect, while shifting
the phase in the $\tilde B$-arm cannot affect $\rho_{\rm fin}$ by
causality. We have also checked independence of $\rho_{\rm fin}$ on
$\varphi_{2}$ and $\varphi_3$ by explicit calculations. Therefore,
we can choose any values of $\varphi_{2}$ and $\varphi_{3}$. For
convenience, let us choose $\varphi_2=\pi$ and $\varphi_3=0$. Then
using step 2 of the substitution method we obtain
    \begin{eqnarray}
&& a^\dagger = \sqrt{\eta} \, e^{i\varphi_{\rm f}} \tilde a^\dagger
+ \sqrt{1-\eta } \, e^{i\varphi_{\rm f}}  \tilde b^\dagger ,
    \label{a-dagger}\\
&&   b^\dagger = \sqrt{\eta} \, \tilde b^\dagger -  \sqrt{1-\eta }
\, \tilde  a^\dagger ,
    \end{eqnarray}
while step 1 was Eq.\ (\ref{psi-in}). Now substituting
$a^\dagger$ in Eq.\ (\ref{psi-in}) with the expression in Eq.\
(\ref{a-dagger}) (step 3), we obtain
\begin{align}
|\Psi_{\rm out}\rangle & =\sum_{n,k}
\alpha_{n+k} \sqrt{(n+k)!/(n!k!)} \, \eta^{n/2} (1-\eta)^{k/2}
    \notag\\
&\times e^{i (n+k)\varphi_{\rm f}}  |n\rangle|k\rangle ,
     \label{Psi-out}\end{align}
where in the notation $|n\rangle|k\rangle=[(\tilde a^\dagger)^n (\tilde
b^\dagger)^k /\sqrt{n!k!}\,] \, |{\bf 0}\rangle$ the second state
corresponds to the ancillary second arm (upper arm in Fig.\
\ref{bsp}).

The final state at the receiving resonator can be calculated by
tracing $|\Psi_{\rm out}\rangle\langle \Psi_{\rm out}|$ over the
ancillary state $|k\rangle$, thus obtaining the density matrix
    \begin{align}
 \rho_{\rm fin}& =\sum_{j,n,m}\alpha_{n+j}\alpha^{*}_{m+j} \sqrt{(n+j)!(m+j)!} /(j!\sqrt{n!m!})\notag\\
& \hspace{0.8cm} \times\eta^{(n+m)/2}
 (1-\eta)^je^{i(n-m)\varphi_{\rm f}}) \,|n\rangle\langle m|, \quad
    \label{rho-fin-ap}\end{align}
 where the sums over $j$, $n$, and $m$ are all from 0 to $\infty$.
Note that this result has been derived for a pure initial state
(\ref{psi-in}) in the emitting resonator. However, it is easy to
generalize Eq.\ (\ref{rho-fin-ap}) to an arbitrary initial state
$\rho_{\rm in}$ by replacing $\alpha_{n+j}\alpha^{*}_{m+j}$ with
$(\rho_{\rm in})_{n+j,m+j}$.

To find the fidelity of the quantum state transfer for the initial state
(\ref{psi-in}), we calculate the overlap $\langle \psi_{\rm in} |
\rho_{\rm fin} |\psi_{\rm in}\rangle$, thus obtaining
    \begin{align}
F_{\rm st}&=\sum_{j,n,m}\frac{\sqrt{(n+j)!(m+j)!}}{j!\sqrt{n!m!}} \,
\alpha_{n}^{*}\alpha_{m}\alpha_{n+j}\alpha^{*}_{m+j}
    \notag\\
&\hspace{0.9cm} \times \eta^{(n+m)/2}(1-\eta)^j
e^{i(n-m)\varphi_{\rm f}},
    \label{F-st-gen-2}\end{align}
which is Eq.\ (\ref{F-st-gen}) in the main text. For a mixed input
state $\rho_{\rm in}$ we can find the resulting state $\rho_{\rm
fin}$ as discussed above and then use the Uhlmann fidelity
definition \cite{N-C-book} $F_{\rm st}=[\mbox{Tr} \sqrt{\sqrt{\rho_{\rm in}}\,
\rho_{\rm fin} \sqrt{\rho_{\rm in}}} ]^2$.

If instead of an arbitrary state (\ref{psi-in}) we transfer a qubit
state $|\psi_{\rm in}\rangle = \alpha_{0}|0\rangle +\alpha_1
|1\rangle$, then in Eq.\ (\ref{Psi-out}) there are only three terms
because $\alpha_{n+k}=0$ if $n+k>1$. This reduces Eq.\
(\ref{Psi-out}) to Eq.\ (\ref{quantum-qubit}) in the main text.
Similarly, Eq.\ (\ref{rho-fin-ap}) reduces to Eq.\
(\ref{quantum-qubit-rho}) and Eq.\ (\ref{F-st-gen-2}) reduces to
    \be
    F_{\rm st} = |\alpha_0|^4 + \eta |\alpha_1|^2
    +|\alpha_0|^2|\alpha_1|^2 (1-\eta+2\sqrt{\eta}\cos \varphi_{\rm f}) .
    \ee
To average this fidelity over the Bloch sphere of the initial state,
we can either average it over 6 points at the ends of the three axes
($\pm$X, $\pm$Y, $\pm$Z) or use the averaging formulas
$\overline{|\alpha_0|^4}=\overline{|\alpha_1|^4}=1/3$,
$\overline{|\alpha_0|^2 |\alpha_1|^2}=1/6$, thus obtaining average
state fidelity
    \be
\overline{F}_{\rm st} =\frac{3+\eta+2\sqrt{\eta}\cos{\varphi_{\rm
f}}}{6},
    \label{F-st-aver-1}\ee
which can be converted into the process fidelity $F_{\chi}$ using the standard rule, $F_{\chi}=1-(3/2)(1-\overline{F}_{\rm st})$.

\subsection{Decrease of the average state fidelity due to photons
in the environment}

So far we have assumed the initial state of the receiving
resonator and all environmental modes in Eq.\
(\ref{Bf-general}) to be vacuum. A natural question is what happens when there
are some photons in the environment (including the initial state of
the receiving resonator). In particular, it is interesting to determine whether the
average fidelity $\overline{F}_{\rm st}$ of the qubit state transfer
can increase, or always decreases. Below we show that the average
fidelity always decreases due to a non-vacuum state of the
environment.

We consider a simplified model, in which the main input of the beam
splitter in Fig.\ \ref{bsp} is in a qubit state $|\psi_{\rm
in}\rangle = \alpha_{0}|0\rangle +\alpha_1 |1\rangle$, while the
second input (modeling the environment) is in an arbitrary state, so
that the total state is
    \be
|\Psi_{\rm in}\rangle = (\alpha_{0}|0\rangle +\alpha_1
|1\rangle)\sum_n \beta_n |n\rangle ,
    \label{Psi-in-app}\ee
where $|\alpha_0|^2+|\alpha_1|^2=1$ and $\sum_n |\beta_n|^2=1$.
 Neglecting for simplicity the transfer phase, $\varphi_{\rm
f}=0$, choosing the other phases as $\varphi_2=\pi$ and
$\varphi_{3}=0$,  and using the substitution method described
above, we find the output state
\begin{align}\label{in}
|\Psi_{\rm out}\rangle &=\sum_{k,m}\frac{\sqrt{(k+m)!}}{\sqrt{k!m!}}
\, \beta_{k+m}(-\sqrt{1-\eta})^m(\sqrt{\eta})^{k}
    \notag\\
&\times\Big[\alpha_{0}|m\rangle|k\rangle+\alpha_{1}\sqrt{\eta}\,
 \sqrt{m+1} \, |m+1\rangle|k\rangle
    \notag\\
&\hspace{0.5cm} + \alpha_{1}\sqrt{1-\eta}\, \sqrt{k+1}\, |m\rangle
|k+1\rangle\Big]. \,\,\,
\end{align}
We then trace over the ancillary arm state to find the resulting
density matrix $\rho_{\rm fin}$, which can now contain non-zero
elements $(\rho_{\rm fin})_{mn}$ for arbitrary $m$ and $n$. However,
the state fidelity for the qubit transfer depends only on the
elements within the qubit subspace, $ F_{\rm st} =
|\alpha_{0}|^2(\rho_{\rm fin})_{00} + |\alpha_{1}|^2(\rho_{\rm
fin})_{11} +2\,\mbox{Re}[\alpha_{0}^{*}\alpha_{1}(\rho_{\rm
fin})_{01}]$. Averaging $F_{\rm st}$ over the initial qubit state
\cite{Nie02,Hor99,Kea12}, we obtain after some algebra
\begin{eqnarray}\label{Fid}
&& \hspace{-0.5cm} \overline{F}_{\rm
st}=\frac{1}{6}(3+\eta+2\sqrt{\eta}) -\sum_{n=1}^\infty C_{n}(\eta)
\, |\beta_{n}|^2 ,
    \\
&& \hspace{-0.5cm} C_{n}(\eta)=\frac{1}{6}\big\{(3+\eta+2\sqrt{\eta})(1-\eta^{n})\notag\\
&&\hspace{-0.2cm}
 +n(1-\eta)\eta^{n-1}[2\eta+2\sqrt{\eta}-(1-\eta)(2n+1)]\big\}.\quad
 \,\,\,
    \label{cj}.
\end{eqnarray}
The first term in Eq.\ \eqref{Fid} is the average fidelity when
there are no photons in the environment [see Eq.\
(\ref{F-st-aver-1}) with $\varphi_{\rm f}=0$], while the second term
is due to the environmental photons ($|\beta_{n}|^2$ is the
probability of having $n$ photons). We numerically checked that the
coefficients $C_{n}(\eta)$ are always positive for $n\geq 1$ and
$\eta \in [0,1]$. Therefore, the presence of photons in the
environment always decreases the average fidelity of a qubit
transfer. Note that Eq.\ (\ref{Fid}) does not depend on the choice
of $\varphi_2$ and $\varphi_3$, since these phases can be produced
by phase shifters in the ancillary $B$-arm and $\tilde B$-arm in
Fig.\ \ref{bsp}. The phase shifter in the $\tilde B$-arm cannot
affect $\rho_{\rm fin}$, while the phase shifter in the $B$-arm
changes only the phase of the ancillary input state and therefore
does not change $|\beta_n|^2$ in Eq.\ (\ref{Fid}).

In the case when $\eta \approx 1$, we can approximate Eq.\
(\ref{cj}) as $C_n(\eta)\approx (5/3)(1-\eta )\, n$. The average
fidelity is then
    \be
\overline{F}_{\rm st} \approx 1 - \frac{1-\eta}{3} -\frac{5}{3}\,
(1-\eta)\, \overline{n}_{\rm e}, \,\,
    \label{F-st-aver-noise}\ee
where $\overline{n}_{\rm e}=\sum_n n |\beta_n|^2$ is the average
number of photons in the environmental mode. Note that the effect of
non-zero $\overline{n}_{\rm e}$ is suppressed at $1-\eta \ll 1$.
Equation (\ref{F-st-aver-noise}) can be used for an estimate of the
effect of finite temperature. However, we emphasize that modeling of
the environmental noise with a single beam splitter is an
oversimplification, so Eq.\ (\ref{F-st-aver-noise}) gives a
qualitative description, but is not intended to accurately describe
the effect of environmental noise on the quantum state transfer
protocol.

\section{Tunable coupler theory}

In this Appendix  we consider the tunable coupler realized
experimentally in Refs.\ \cite{Yin13,Wen14}, and derive formulas for
the transmission and refection amplitudes $\textbf{t}$ and
$\textbf{r}^{\rm in}$ used in Sec.\ \ref{sec-variable-detuning}. We
also discuss the change of the resonator frequency due to the changing
complex phase of $\textbf{r}^{\rm in}$. Since the theory is the same
for both resonators, we omit the resonator index, assuming, e.g.,
the receiving resonator. The discussion in this Appendix follows the
discussion in Sec.\ III of the Supplementary Information of Ref.\
\cite{Yin13}.

There will be a difference in the choice of rotating frame between the
main text and this Appendix. In the main text we use the rotating
frame $e^{-i\omega t}$, which is standard in optics. However, in
this Appendix we will need a language of impedances, which
traditionally assumes the rotating frame $e^{i\omega t}$. Therefore,
we will have to derive formulas for $\textbf{t}$ and $\textbf{r}$ in
the rotating frame  $e^{i\omega t}$, and then we will need to
conjugate the final results to convert them into for $\textbf{t}$
and $\textbf{r}$ for the rotating frame  $e^{-i\omega t}$.

\begin{figure}[t]
\includegraphics[width=8cm]{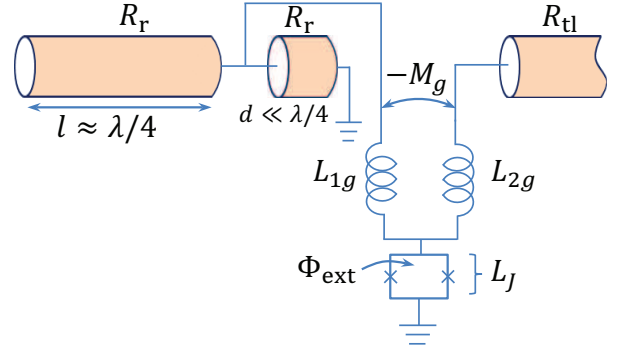}
   \caption{Schematic of the tunable coupler of Refs.\
\cite{Yin13,Wen14} between the $\lambda/4$ microwave resonator (at
the left) and the transmission line (at the right). A voltage taken
at the distance $d$ from the resonator end is applied to a
transformer with a negative mutual inductance $-M_g$ and a SQUID
providing positive Josephson inductance $L_J$. External flux
$\Phi_{\rm ext}$ controls $L_J$, thus controlling the effective
mutual inductance $M=-M_g+L_J$. The wave impedances of the lines are
$R_{\rm r}$ and $R_{\rm tl}$.
   } \label{tuna}
\end{figure}

The schematic of the tunable coupler is shown in Fig.~\ref{tuna}. A quarter-wavelength ($\lambda/4$) microwave resonator is divided into two unequal parts, and the voltage signal for the coupler is taken at the distance $d$ ($d\ll \lambda/4$) from the end, which is shorted to the ground, while the other end is terminated with a break so that the total length is $l+d\approx \lambda/4$. The coupler consists of a transformer with geometrical inductances $L_{1g}$ and $L_{2g}$ and negative mutual inductance $-M_{g}$, which is in series with a dc SQUID, providing a positive Josephson inductance $L_J$. This inductance is controlled by an external magnetic flux $\Phi_{\rm ext}$, $L_{J}=\Phi_{0}/[2\pi\sqrt{I_{c1}^2+I_{c2}^2+2I_{c1}I_{c2}\cos(2\pi \Phi_{\rm ext}/\Phi_{0})}]$, where $\Phi_{0}=h/2e$ is the magnetic flux quantum and $I_{c1}$, $I_{c2}$ are the critical currents of two Josephson junctions, forming the SQUID. Thus the external flux controls the total mutual inductance $M=-M_g+L_J$, which determines the coupling between the resonator and transmission line; in particular, there is no coupling when $M=0$. Note that the wave impedance $R_{\rm r}$ of the resonator may be different from the impedance $R_{\rm tl}$ of the transmission line.

\begin{figure}[t]
\includegraphics[width=8cm]{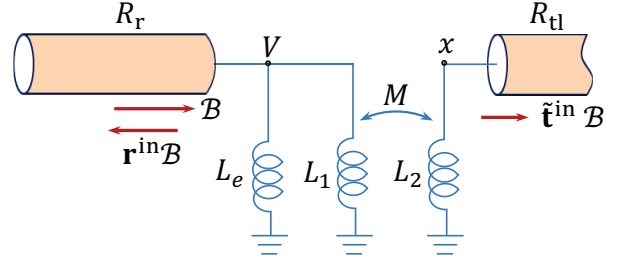}
  \caption{The simplified schematic of Fig.\ \ref{tuna}, with the
$d$-long piece of the resonator replaced by inductance $L_e$, and
the transformer in series with SQUID replaced by an effective
transformer with mutual inductance $M$. An incident wave with
voltage amplitude $\mathcal{B}$ creates voltages $V$ and $x$ across
the inductors $L_{1}$ and $L_{2}$. The wave is reflected as
$\textbf{r}^{\rm in}\mathcal{B}$ and transmitted as $\tilde{\bf
t}^{\rm in} \mathcal{B}$ (the superscript ``in'' indicates the wave
coming from inside the resonator and the tilde sign indicates the
actual transmission amplitude, as opposed to the effective amplitude
${\textbf t}$). In our case ${\textbf r}^{\rm in}\approx -1$ and  $|
\tilde{\bf t}^{\rm in}| \ll 1$.
  } \label{tunb}
\end{figure}

For the analysis let us first reduce the schematic of Fig.\ \ref{tuna} to the schematic of Fig.\ \ref{tunb} by replacing the $d$-long part of the resonator with an effective inductance $L_e$ and also replacing the transformer and SQUID with an effective transformer with inductances $L_1$, $L_2$, and mutual inductance $M$,
\begin{equation}\label{LL}
  L_{1}=L_{1g}+L_{J}, \,\,\, L_{2}=L_{2g}+L_{J}, \,\,\, M=-M_{g}+L_{J}.
\end{equation}
We emphasize that $M$ can be both positive and negative, so the coupling changes sign when $M$ crosses zero (the coupler is OFF when $M=0$). Note that by varying $M$ we also slightly change $L_{1}$ and $L_{2}$,
\begin{equation}\label{gg}
  L_{1}=L_{1g}+M_{g}+M, ~~ L_{2}=L_{2g}+M_{g}+M.
\end{equation}

It is easy to calculate the effective inductance $L_e$. If there is no coupler ($L_1=\infty$) and a voltage wave $\mathcal{B}e^{i\omega t}$  comes from the resonator side (from the left in Fig.\ \ref{tuna}), then it is reflected as $-\mathcal{B}e^{i \omega t}$, and the voltage at a distance $d$ is then $V=\mathcal{B}e^{i\omega t}[\exp(i\omega d/v)-\exp(-i\omega d/v)]=2i\mathcal{B} e^{i\omega t}\sin(\omega d/v)$, where $v$ is the speed of light in the resonator. The current (to the right) at this point is $I=(\mathcal{B}/R_{\rm r})e^{i\omega t}[\exp(i\omega d/v)+\exp(-i\omega d/v)]=2(\mathcal{B}/R_{\rm r})\cos(\omega d/v)$. Therefore, the wave impedance is $Z=V/I=iR_{\rm r}\tan(\omega d/v)$, which is the same, $Z=i\omega L_{e}$, as for an inductance
\begin{equation}\label{Le-ind}
  L_{e}=\frac{R_{\rm r}}{\omega}\tan \frac{\omega d}{v}=\frac{R_{\rm r}}{\omega}\tan \frac{2\pi d}{\lambda}.
\end{equation}

Next, let us calculate the transmission and reflection amplitudes
$\tilde{\bf t}^{\rm in}$ and ${\textbf r}^{\rm in}$ for the
effective circuit shown in Fig.\ \ref{tunb}. (Here the superscript
``in'' reminds us that the wave is incident from inside of the
resonator, and the tilde sign in $\tilde{\bf t}^{\rm in}$  means
that we consider the actual transmission amplitude, which is
different from the effective amplitude $\textbf t$). Assume that a
voltage wave with amplitude $\mathcal{B}$ is incident onto the
coupler from the resonator (we omit the exponential factor
$e^{i\omega t}$). The wave is
reflected as $\textbf r^{\rm in}\mathcal{B}$ and transmitted as
$\tilde{\bf t}^{\rm in} \mathcal{B}$. For a weak coupling, which we
consider in this paper, $\textbf r^{\rm in}\approx-1$ and
$|\tilde{\bf t}^{\rm in}|\ll 1$. The voltage across $L_{1}$ is
$V=(1+{\textbf r}^{\rm in})\mathcal{B}$, while the voltage across
$L_{2}$ is denoted by $x$. The current flowing into $L_1$ is
$I_{1}=(1-{\textbf r}^{\rm in})\mathcal{B}/R_{\rm r}-V/(i\omega
L_e)$, while the current flowing (down) into $L_{2}$ is
$I_{2}=-x/R_{\rm tl}$. Using the currents $I_{1}$ and $I_{2}$, we
write transformer equations for voltages $x$ and $V$ as
\begin{eqnarray}\label{x}
  && x=i\omega M\left[\frac{(1- {\textbf r}^{\rm in})\mathcal{B}}{R_{\rm r}}-\frac{(1+{\textbf r})^{\rm in}\mathcal{B}}{i\omega L_{e}}\right]-i\omega L_{2}\frac{x}{R_{\rm tl}}, \,\,\,\, \qquad
    \\
\label{x2}
 && (1+{\textbf r}^{\rm in})\mathcal{B}=i\omega L_{1}\left[\frac{(1-{\textbf r}^{\rm in})\mathcal{B}}{R_{\rm r}}-\frac{(1+ {\textbf r}^{\rm in})\mathcal{B}}{i\omega L_{e}}\right]
   \nonumber\\
 && \hspace{1.9cm} -i\omega M\frac{x}{R_{\rm tl}}. \,\, \qquad
\end{eqnarray}
From these two equations we can find the reflection amplitude
${\textbf r}^{\rm in}$ and the transmission amplitude $\tilde{\bf
t}^{\rm in}=x/\mathcal{B}$ (note that $|\tilde{\bf t}^{\rm
in}|^2R_{\rm r}/R_{\rm tl}+|{\textbf r}^{\rm in}|^2=1$):
\begin{align}\label{rr}
  & {\textbf r}^{\rm in}=-\frac{1-b}{1+b},
   \\
&\tilde{\bf t}^{\rm in}=i\frac{2\omega M}{1+b}\left(\frac{1}{R_{\rm
r}}+\frac{ib}{\omega L_{e}}\right)\frac{1}{1+i\omega L_{2}/R_{\rm
tl}}, \label{tilde-t-in}
\end{align}
where
\begin{eqnarray}
&& b=\frac{\displaystyle \frac{i\omega L_{1}}{R_{\rm
r}}+\frac{\omega^2M^2}{R_{\rm r}R_{\rm tl}(1+i\omega L_{2}/R_{\rm
tl})} }{\displaystyle 1+\frac{L_{1}}{L_{e}}-\frac{i\omega
M^2}{R_{\rm tl}L_{e}(1+i\omega L_{2}/R_{\rm tl})}}
   \label{app-b-1} \\
&& \hspace{0.3cm}   = \frac{i\omega L_{1}/R_{\rm r}}{\displaystyle
\frac{L_1}{L_e}+\left[ 1-\frac{i\omega M^2}{R_{\rm
tl}L_{1}(1+i\omega L_{2}/R_{\rm tl})}\right]^{-1}} . \qquad
  \label{app-b-2}\end{eqnarray}
 Note that the transmission and reflection amplitudes for the wave
incident from outside of the resonator are
    \be
\tilde{\bf t}^{\rm out} = \frac{R_{\rm r}}{R_{\rm tl}} \, \tilde{\bf t}^{\rm in} , \,\,\, {\textbf r}^{\rm out}= - \frac{\tilde{\bf t}^{\rm in}}{(\tilde{\bf t}^{\rm in})^*}\, {\textbf r}^{\rm in}.
    \ee
Since the transmission amplitude depends on the direction, it is convenient to introduce the effective amplitude $\textbf t$, which does not depend on the direction,
    \be
    {\textbf t}= \sqrt{\frac{R_{\rm r}}{R_{\rm tl}}} \, \tilde{\bf t}^{\rm in}= \sqrt{\frac{R_{\rm tl}}{R_{\rm r}}} \, \tilde{\bf t}^{\rm out}, \,\,\,
    | {\textbf t}|^2+|{\textbf r}^{\rm in(out)}|^2=1.
    \label{t-eff}\ee

Equations (\ref{rr})--(\ref{app-b-2}) and (\ref{t-eff}) give us
$\textbf t$ and $\textbf{r}^{\rm}$ in the rotating frame $e^{i\omega
t}$. For the rotating frame $e^{-i\omega t}$ we need to conjugate
$\textbf t$ and $\textbf{r}^{\rm}$ (and $b$), thus obtaining Eqs.\
(\ref{ttn})--(\ref{b-def}) in the main text.

For an estimate let us use the following parameters (similar to the parameters of Ref.\ \cite{Yin13}): $R_{\rm r}=80\,
\Omega$, $R_{\rm tl}=50\, \Omega$,  $L_{1g}=L_{2g}=480$ pH,
$M_g=140$~pH, $\omega /2\pi = 6$ GHz, and $L_e=180$ pH
(corresponding to $d/\lambda=0.013$). Then Eqs.\
(\ref{rr})--(\ref{app-b-2}) and (\ref{t-eff}) for small $M$ give
$b\approx 0.066 i$, ${\textbf r}^{\rm in} \approx - e^{-0.13i}$, and
${\bf t}\approx 0.034 i e^{-0.5i} M/M_g$. The resonator leakage time
is then $\tau \approx (M_g/M)^2 \times 72$ ns.

Note that in the case when $\omega M \ll R_{\rm tl}$, we can replace
the denominator of Eq.\ (\ref{app-b-2}) with $L_1/L_e+1$. Then
    \be
b\approx i\, \frac{\omega L_e /R_{\rm r}}{1+L_e/L_1},
    \label{b-approx}\ee
and if  $\omega L_{e}\ll R_{\rm r}$ (which means $d\ll\lambda/4$),
then $|b|\ll 1$. In this case the reflection and effective
transmission amplitudes (\ref{rr}) and (\ref{t-eff})  can be
approximated (for the rotating frame $e^{i\omega t}$) as
 \begin{eqnarray}\label{rapp}
&& {\textbf r}^{\rm in} \approx -\exp \left[ - \frac{2\omega
L_{e}L_{1}}{R_{\rm r}(L_{1}+L_{e})} \, i\right]
  \\
&& {\textbf t} \approx i\, \frac{2\omega L_{e}M}{\sqrt{R_{\rm r}
R_{\rm tl}} \,(L_{1}+L_{e})}\, \frac{1}{1+i\omega L_2/R_{\rm tl}}.
\qquad
  \label{tapp}\end{eqnarray}
The latter equation shows that in the first approximation the phase
of $\textbf{t}$ does not change with $M$, and for the case $\omega
L_2 \ll R_{\rm tl}$ the value of $\textbf{t}$ is close to being
purely imaginary. Note that Eq.\ (\ref{tapp}) uses the approximation
$1+b\approx 1$ in the denominator of the first factor in Eq.\
(\ref{tilde-t-in}). Without this approximation (still using the
above formula for $b$), the factor $L_1+L_e$ in the denominator of
Eq.\ (\ref{tapp}) should be replaced with a more accurate term
$L_1+L_e+i\omega L_1L_e/R_{\rm r}$. As we checked numerically, this
gives a much better approximation for small $M$ (mostly
for the phase of $\textbf{t}$), but there is no significant
improvement of accuracy for intermediate values of $M$, corresponding
to $|\textbf{t}|\simeq 0.05$.

The resonator frequency $\omega_{\rm r}$ slightly changes when the
mutual inductance $M$ is varied, because this slightly changes the
phase of the reflection amplitude $\textbf{r}^{\rm in}$. The
frequency change can be calculated as
 \begin{equation}\label{ll}
   \delta \omega_{\rm r} \approx 2 \, \omega_0 \,
   \frac{\delta(\arg {\textbf r}^{\rm in})}{2\pi},
    \end{equation}
where the factor of 2 comes from the assumption of a $\lambda /4$
resonator, and as $\omega_0$ we choose the resonator frequency at
$M=0$. [Note the sign difference compared with Eq.\
(\ref{delta-omega}) because of the different rotating frame.]

To estimate the frequency change $\Delta \omega_{\rm r} =\omega_{\rm
r}(M)-\omega_{\rm r}(0)$ to first order, we can expand Eq.\
(\ref{app-b-2}) to linear order in $M$ [which comes from changing
$L_1$  in Eq.\ (\ref{b-approx}) -- see Eq.\ (\ref{gg})] and then use $\delta (\arg {\textbf r}^{\rm
in})=-[2 /(1+|b|^2)]\, \delta |b|$, which follows from Eq.\ (\ref{rr})
for a positive-imaginary $b$. Thus we obtain
    \be
    \Delta \omega_{\rm r} \approx - \frac{\omega_0}{\pi} \,
    \frac{2}{1+ |b|^2}
    \, \frac{\omega_0 L_e^2}{R_{\rm r}(L_1 +L_e)^2}\, M,
    \label{Delta-omega-approx}\ee
where $b$ is given by Eq.\ (\ref{b-approx}), and  $L_1$ should be evaluated at $M=0$. Since $\textbf t$ is also
proportional to $M$ in the first order [see Eq.\
(\ref{tilde-t-in})], the ratio $\Delta \omega_{r}/|\textbf{t}|$ is
approximately constant,
    \be
    \frac{\Delta\omega_{\rm r}}{|\textbf{t}|} \approx -\frac{\omega_0}{\pi}
 \, \frac{\sqrt{1+(\omega_0L_2/R_{\rm tl})^2}}{\sqrt{1+|b|^2}}
  \sqrt{\frac{R_{\rm tl}}{R_{\rm r}}} \,
 \frac{L_e}{L_1+L_e} ,
    \label{Delta-omega-ratio}  \ee
where $L_1$ and $L_2$ should be evaluated at $M=0$, and for typical
experimental parameters $|b|^2$ can be neglected [we keep the very
small terms with $|b|^2$ in Eqs.\ (\ref{Delta-omega-approx}) and
(\ref{Delta-omega-ratio}) to have exact formulas at $M\rightarrow
0$]. This formula describes the numerical dependence $\Delta
\omega_{\rm r}(|\textbf{t}|)$ shown in Fig.\ \ref{Mvst} very well,
giving an exact result at $|{\textbf t}| \rightarrow  0$ and a relative
deviation of 3.2\% at $|{\textbf t}| =0.1$. It is interesting that
the dependences of $|\textbf{t}|$ and $\Delta \omega_{\rm r}$ on $M$
are both significantly nonlinear (see, e.g., the dashed line in
Fig.\ \ref{Mvst}); however, these nonlinearities partially
compensate each other to produce a smaller nonlinearity in $\Delta
\omega_{\rm r}(|\textbf{t}|)$.

While Eq.\ (\ref{Delta-omega-approx}) gives only the linear component of the dependence $\Delta \omega_{\rm r}(M)$, a better approximation can be based on using Eq.\ (\ref{b-approx}) to find $b(M)-b(0)$ and then convert it into $\Delta \omega_{\rm r}$ via Eq.\ (\ref{ll}). In this way we obtain
    \be
    \Delta \omega_{\rm r} \approx -
    \, \frac{2\omega_0^2 L_e^2/(1+ |b|^2)}{\pi R_{\rm r}(L_{1g}+M_g +L_e)(L_{1g}+M_g+L_e+M)}\, M,
    \label{Delta-omega-approx-2}\ee
in which the term $|b|^2$ can be neglected. This formula gives a nonlinear dependence $\Delta \omega_{\rm r}(M)$ due to the presence of $M$ in the denominator. We checked that this formula correctly describes about 80\% of the numerical nonlinearity of the $\Delta \omega_{\rm r}(M)$ dependence for the parameters of Fig.\ \ref{Mvst}. There is a similar dependence on $M$ in the denominator of Eq.\ (\ref{tapp}) for $\textbf{t}(M)$ dependence, thus explaining why the two nonlinearities partially cancel each other to produce a much more linear dependence $\Delta
\omega_{\rm r}(|\textbf{t}|)$ in Fig.\ \ref{Mvst}.

\end{document}